%% file: main.tex
\documentclass[letterpaper,twocolumn,10pt]{article}
\usepackage{times}

\usepackage{usenix2019_v3}
\usepackage{tikz}
\usepackage{threeparttable}
\usepackage{soul}
\usepackage{multirow}
\usepackage{tabularx}
\newcolumntype{Y}{>{\centering\arraybackslash}X}
\usepackage{amsmath,pifont, mathtools}
\usepackage{amsmath}
\usepackage[overload]{empheq}
\usepackage{nccmath}
\usepackage{algorithm}
\usepackage{cases}
\usepackage{bbding}
\usepackage{graphicx}
\usepackage{pifont}
\usepackage{wasysym}
\usepackage{algorithmicx}
\usepackage[noend]{algpseudocode}
\usepackage{mathtools}
\usepackage{float}
\usepackage{subcaption}
\usepackage{adjustbox}
\usepackage{makecell}%
\usepackage{tikz}
\makeatletter
\newcommand*{\rom}[1]{\expandafter\@slowromancap\romannumeral #1@}
\makeatother

\usepackage[english]{babel} 
\usepackage{amssymb}
\newcommand{\cmark}{\ding{51}}
\newcommand{\xmark}{\ding{55}}

\usepackage{booktabs}
\usepackage{hyperref}

\usepackage{cleveref}
\crefname{section}{}{\S\S}
\crefdefaultlabelformat{\S#2#1#3}

\usepackage{xspace}
\usepackage{comment}
\usepackage[normalem]{ulem}
\usepackage{enumitem}
\setlist{nolistsep}

\usepackage{listings}
\usepackage{color}
\usepackage{caption}
\usepackage{textcomp}
\usepackage{url}

\usepackage{breakurl}
\definecolor{dkgreen}{rgb}{0,0.6,0}
\definecolor{gray}{rgb}{0.5,0.5,0.5}
\definecolor{mauve}{rgb}{0.58,0,0.82}
\DeclareCaptionLabelFormat{andtable}{#1~#2  \&  \tablename~\thetable}
\usepackage{balance}

\newcommand{\revise}[1]{{\color{black} #1}}
\newcommand{\sys}{Lina\xspace}
\newcommand{\op}{all-to-all\xspace}
\newcommand{\Op}{All-to-all\xspace}
\newcommand{\ar}{allreduce\xspace}

\graphicspath{{figures/}}
\hypersetup{final}
\begin{document}

\pagestyle{empty}
\title{Accelerating Distributed MoE Training and Inference with \sys}
\date{}
\author{
{\rm Jiamin Li}\\
City University of Hong Kong
\and
{\rm Yimin Jiang}\\
ByteDance Inc.
\and
{\rm Yibo Zhu}\\
Unaffiliated
\and
{\rm Cong Wang}\\
City University of Hong Kong
\and
{\rm Hong Xu}\\
The Chinese University of Hong Kong
} 
\maketitle
\input{abstract}

\input{introduction}
\input{motivation}
\input{design}
\input{training}

\input{inference}
\input{implementation}
\input{evaluation}
\input{discussion}
\input{related}

\section{Conclusion}
We presented \sys, a new system that accelerates \op in distributed MoE. 
Through a systematic analysis, we build \sys upon two key ideas: first to prioritize \op over \ar using tensor partitioning and pipelining to improve its bandwidth in training, and second to dynamically balance the workload with token-level expert selection pattern in inference.
We implemented \sys over DeepSpeed and performed extensive testbed evaluation using A100 GPUs and 100Gbps InfiniBand to show that \sys significantly improves training efficiency and inference time. 

\section*{Acknowledgment}
We thank the anonymous ATC’23 reviewers and our
shepherd Myoungsoo Jung for their constructive and valuable
comments. We also thank the anonymous reviewers from
NSDI’22 for their feedback
that helped improve the paper. This work was supported in
part by funding from the Research Grants Council of Hong
Kong (N\_CityU139/21, C2004-21GF, C7004-22G, R1012-21, R6021-20F,
and 11209520), and from CUHK (4055138).
 
\bibliographystyle{plain}
\bibliography{main}
\clearpage
\end{document}

%% file: abstract.tex
%!TEX root = main.tex
\begin{abstract}

\end{abstract}

%% file: introduction.tex
%!TEX root = main.tex

\section{Introduction}
\label{sec:introduction}
Recent advances in deep learning have shown that a model's quality typically improves with more parameters~\cite{gpt-3, devlin2018bert, vaswani2017attention, fedus2021switch, kaplan2020scaling}. 
Many new frontiers in Computer Vision (CV) and Natural Language Processing (NLP) have been explored using large dense models \cite{shazeer2017outrageously, du2021glam, msMoE}. 
While effective in terms of model quality, the computation cost of model training and serving is extremely high. ChatGPT~\cite{ChatGPT}, an impressive chatbot released by OpenAI, is estimated to spend 3 million dollars per month to serve user requests. Wider adoption and development of these models are impeded by the exorbitant compute cost.

Following the basic idea of curbing the computation cost of massive models, \textit{sparsely activated} models have recently been introduced~\cite{bengio2015conditional, shazeer2017outrageously, fedus2021switch, liang2022mvit}. The \textit{Mixture-of-Experts} (MoE) structure is now one of the most popular ways to implement sparse activation~\cite{shazeer2017outrageously, bengio2015conditional, bengio2013estimating, zuo2021taming}. 
For each input, instead of using all parameters, an MoE model selects just a few of them, i.e. \textit{experts}, for processing. 
This leads to sub-linear scaling of FLOPs needed with model size.
Recent literature~\cite{artetxe2021efficient, zoph2022designing, du2021glam, msMoE, jia2020whale, komatsuzaki2023sparse} has proven the potential of MoE models. For instance, Google develops a family of language models named GLaM using MoE~\cite{du2021glam}. Compared to GPT-3 with 175 billion parameters, the largest GLaM has 1.2 trillion parameters while only consuming 1/3 of the energy for training. Meanwhile, GLaM still achieves better zero-shot and one-shot performance than GPT-3.
Microsoft reports that their MoE-based language models achieve a 5x training cost reduction compared to a dense model with the same model quality~\cite{msMoE}. 

Given the uptake of MoE, there have been several systems for efficient
MoE training and inference, including Google's Mesh
TensorFlow~\cite{shazeer2018mesh}, Meta's FairScale~\cite{FairScale2021},
Microsoft's DeepSpeed~\cite{deepspeed} and Tutel~\cite{tutel}, etc. They provide
APIs for users to replace the conventional dense layers with MoE layers with
minimal code changes. They adopt both data parallelism and expert parallelism to
accelerate the training and inference. That is, each device (e.g.
GPU) is assigned with a unique expert, and uses \op to receive
inputs from other devices and then sends the gradients back to them accordingly.
During training, \ar is then used to aggregate non-expert gradients in the backward pass. 

We focus on the efficiency of distributed MoE training and inference in this work. 
As some~\cite{rajbhandari2022deepspeedmoe, lepikhin2020gshard,he2022fastermoe} has shown, the \op operation is the main bottleneck. 
\Op blocks the subsequent computation operations and needs to be invoked two times in the forward pass and another two in the backward pass for each MoE layer. 
Interestingly, the main causes for \op being the bottleneck are different in training and inference. In training, \op and \ar often contend for network bandwidth when they overlap in the backward pass, leading to a prolonged blocking period to the computation. 
Inference, on the other hand, presents a highly-skewed expert popularity driven by real-world requests. 
Devices with popular experts have to handle much more data than others. Not only does it delay the launch of \op, but it also causes imbalanced transfer size and bandwidth utilization across the devices, both of which are detrimental.

We are thus motivated to systematically tackle the \op bottleneck. 
Our solution is \sys, a system that accelerates both MoE training and inference. 

\emph{In training}, we prioritize \op over \ar in order to improve its bandwidth. Existing MoE systems launch separate CUDA streams for the expert-parallel and data-parallel process groups which correspond to \op for expert and \ar for non-expert parameters, respectively. As there is no coordination between these streams, \op and \ar can overlap and fair-share the network bandwidth.  Unlike \ar, \op is blocking and cannot be made parallel with the computation process. Thus, prioritizing \op in the backward pass and avoid concurrent \ar is crucial to reducing the blocking period. 

To efficiently prioritize \op, we adopt tensor partitioning which breaks down a tensor into smaller chunks, each of which forms a {micro-op}.
With micro-ops, simple priority scheduling can be applied to guarantee full bandwidth for \op while allowing \ar micro-ops to make progress when \op is not present. 
In addition, micro-ops allow the expert computation to be pipelined with \op. 

\emph{In inference}, we dynamically schedule the resources for each expert in order to balance the workload of each device, thereby alleviating the imbalanced \op transfer size and bandwidth. 
Intuitively popular experts should be given more resources while the rest may be served with less resources.
% in MoE layer often receive and process more data while the others waiting are idle during this period. 
The key challenge here is to efficiently and accurately obtain the expert popularity \textit{before} the selection is actually done by the gating network, for every batch of input at each MoE layer, so scheduling benefit can be maximized with minimal overheads. 
% It is challenging to deploy the model with a ideal setting as the expert selection is unknown until the requests arrive. 
Fortunately, we find the experts selected by each token across the layers demonstrate clear patterns, which allow us to estimate the expert distribution of the upcoming layer based on the past selection results from the preceding layers.
%  collected during training. 
We adopt a two-phase scheduling approach that fine-tunes the estimation based allocation only when the actual expert popularity deviates too far.

We build \sys based on DeepSpeed MoE \cite{deepspeed} and PyTorch, and evaluate it on a 
cluster with up to 16 Ampere A100 GPUs with 40GB memory and 100Gbps InfiniBand. 
Results show that \sys accelerates \op by at least 2.21x, and achieves on average 1.57x speedup in overall training step time compared to state-of-the-art system DeepSpeed. The median and 95\%ile inference time is reduced by 1.45x and 1.63x.

Our contributions can be summarized as follows:
\begin{itemize}[leftmargin=*]
    \item We present an in-depth empirical analysis of distributed MoE to show the main causes for \op to be the performance bottleneck in training and inference.
    \item We propose to prioritize \op over \ar in order to improve its bandwidth and reduce its blocking period in distributed training. \sys's scheduler incorporates tensor partitioning and pipelining to perform micro-op scheduling.
    \item We examine the pattern in expert selection of MoE layer and propose to estimate the expert popularity to conduct resource scheduling in advance during inference. \sys adopts a two-phase scheduling scheme to minimize the overhead.
    \item We implement a concrete prototype system and conduct comprehensive testbed experiments to demonstrate the benefits of our design in a realistic GPU cluster setting. 
\end{itemize}

%% file: motivation.tex
%!TEX root = main.tex
\section{Background and Motivation}
\label{sec:motivation}
We start with an introduction on MoE and a widely-adopted distributed system for MoE model in~\Cref{subsec:background}. Then, we motivate our idea by analyzing the performance bottleneck (i.e. \op) in distributed MoE training and inference in~\Cref{subsec:bottleneck}.

\subsection{A Primer on MoE}
\label{subsec:background}
Mixture-of-Experts (MoE) has been adapted to different types of DNN models
, and exhibits great potential in improving the performance of language models in particular. GShard~\cite{lepikhin2020gshard} and Switch Transformer~\cite{fedus2021switch} are two seminal works on scaling Transformer-based language models with MoE layers. We focus on MoE in Transformer-based models in this work.

\begin{figure}[t]
    \begin{subfigure}{0.20\textwidth}
        \centering
        \includegraphics[width=0.6\linewidth]{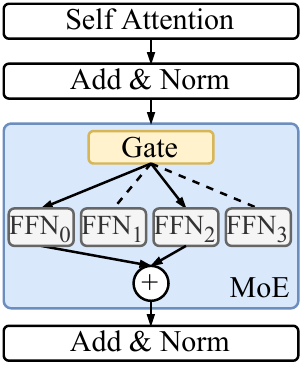}
        \caption{There are four experts and the gate selects two experts.}
        \label{fig:moe_def}
    \end{subfigure}%
    \hfill
    \begin{subfigure}{0.25\textwidth}
        \centering
        \includegraphics[width=0.9\linewidth]{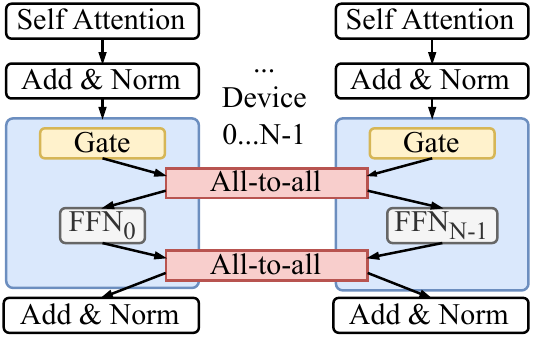}
        \caption{Distributed MoE. Data parallelism and expert parallelism 
        are used.}
        \label{fig:moe_distributed}
    \end{subfigure}
    \vspace{-3mm}
    \caption{MoE layer in Transformer-based models.}
    \vspace{-3mm}
\end{figure}

Transformer-based models normally use an MoE layer to replace the feed-forward network (FFN) layer. An MoE layer consists of multiple FFNs each serving as an \textit{expert}, and a gating network (Figure~\ref{fig:moe_def}). Every expert is a fully-connected two-layer network using ReLU activation but with different parameters. The gating network takes in the embedding vector of each token and multiplies them with its trainable matrix. Based on the results, it dispatches the token to a small number of experts (usually one or two). The final output of the MoE layer is the weighted sum of outputs from the selected expert(s). The sparsity nature of MoE improves the model scaling in size without increasing the training cost and naturally leads to a dynamically-changing model graph.

\revise{
\noindent\textbf{Load balancing loss.} In MoE training, an auxiliary loss is introduced to evaluate the token distribution among the experts~\cite{fedus2021switch}. The objective is to achieve a uniform distribution of tokens across the experts, thereby preventing an excessive concentration of tokens in a single expert. By minimizing this loss term, we encourage but do not enforce the gating network to produce a perfectly balanced token distribution.

The standard practice is to calculate the auxiliary loss of each MoE layer and sum them with the training loss using an appropriate weight. Previous research has demonstrated the effectiveness of this approach~\cite{chen2023sparse, liang2022mvit, shen2022se}. However, it should be noted that achieving a perfectly balanced distribution, where the auxiliary loss converges to zero, is challenging~\cite{zhou2022mixture, chi2022representation}.
During MoE inference, the trained gating network is utilized to dispatch tokens to the experts based on their respective embeddings. This process is solely driven by the characteristics of the token embeddings.
}

\noindent\textbf{Hybrid parallelism in distributed MoE.} 
Training and serving MoE models in a
distributed manner are necessary due to the tremendous compute requirement of
large-scale language models~\cite{MLSYS2022_98dce83d}. For efficiency, both data parallelism and
MoE-specific \textit{expert parallelism} (as a form of model parallelism) are applied~\cite{lepikhin2020gshard,fedus2021switch}.
Existing MoE systems \cite{lepikhin2020gshard,fedus2021switch,shazeer2018mesh,FairScale2021,deepspeed,tutel} allocate one unique compute device (e.g., GPU) for each expert in expert parallelism.
An \op communication is then needed to send tokens to their experts selected by the gating network, and another \op is needed to send tokens back to the device they belong to in data
parallelism to finish the rest of the forward pass as shown in Figure~\ref{fig:moe_distributed}. 

\subsection{Bottleneck Analysis}
\label{subsec:bottleneck}
Much prior work has identified that the introduction of \op in MoE causes performance inefficiency in Transformer-based models
\cite{tutel,rajbhandari2022deepspeedmoe, zoph2022designing}. We extract the completion time of \op operations in both training and inference in our GPU cluster as shown in Table~\ref{table:alltoall_to_step}. 
All our experiments in this section use the same testbed and settings.
Overall, \op takes an average of 34.1\% --- a significant fraction of the step time. Interestingly, though the bottleneck brought by \op is universal in both MoE training and inference, the causes differ. In the following, we motivate our work by analyzing how \op affects the efficiency of training and inference, respectively.
\begin{table}[t]   
    \centering
    \resizebox{0.9\columnwidth}{!}{
        \begin{tabular}{@{}cccccc@{}}
            \toprule
            \multicolumn{1}{c}{\# Experts} & \multicolumn{1}{c}{Model}              & \multicolumn{2}{c}{Training (ms)}                                 & \multicolumn{2}{c}{Inference (ms)}                                 \\ \cmidrule(l){3-6} 
            \multicolumn{1}{c}{GPUs}       & \multicolumn{1}{c}{\#Layers \& Params} & \multicolumn{1}{c}{\Op} & \multicolumn{1}{c}{Ratio} & \multicolumn{1}{c}{\Op} & \multicolumn{1}{c}{Ratio} \\ \midrule
                & 12L + 117M     & 259     & 36.7\%   &  73    &   27.4\%  \\
            4   & 24L + 233M     & 589     & 35.4\%   &  103   &   26.2\%     \\
                & 36L + 349M     & 979     & 38.2\%   &  153   &   28.3\% \\ \midrule
                & 12L + 419M     & 333     & 39.5\%   &  102   &   32.5\% \\
            16  & 24L + 838M     & 715     & 37.6\%   &  177   &   31.7\% \\
                & 36L + 1.2B     & 1145    & 36.8\%   &  243   &   27.4\% \\ \bottomrule
            \end{tabular}}
    \captionof{table}{The completion time of \op and its ratio in training and inference task of Transformer-XL~\cite{dai2019transformer} in different number of experts per layer. Training and inference have the same batch size here. Each FFN layer is replaced with MoE and the number of experts is equal to the number of GPUs similar to the common practice~\cite{fedus2021switch}. A100 GPUs with 40GB memory and 100Gb/s InfiniBand are used. We use the MoE implementation in DeepSpeed.
    }
    \vspace{-4mm}
    \label{table:alltoall_to_step}
\end{table}

\begin{figure}[t]
    \centering
    \includegraphics[width=0.85\linewidth]{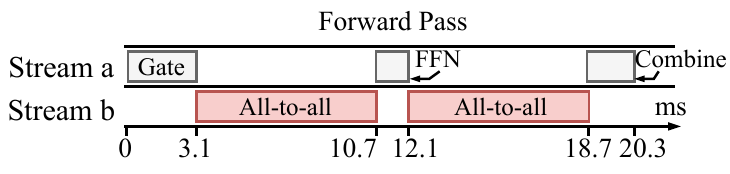}
    \vspace{-3mm}
    \caption{Timeline of forward pass an MoE layer. 
    We simplify the presentation by bundling GPU kernels here:  
    The computation kernels are grouped by their roles in the MoE layer 
    into Gate, FFN and Combine. The Combine operation 
    involves reshaping the tensors and computing the weighted output. 
    The timeline is taken from a sample run of the 419M-parameter model in Table~\ref{table:alltoall_to_step}.}
    \vspace{-4mm}
    \label{fig:moe_timeline}
\end{figure}

\begin{figure}[t]
    
    \begin{minipage}{0.23\textwidth}
        \centering
        \includegraphics[width=0.95\linewidth]{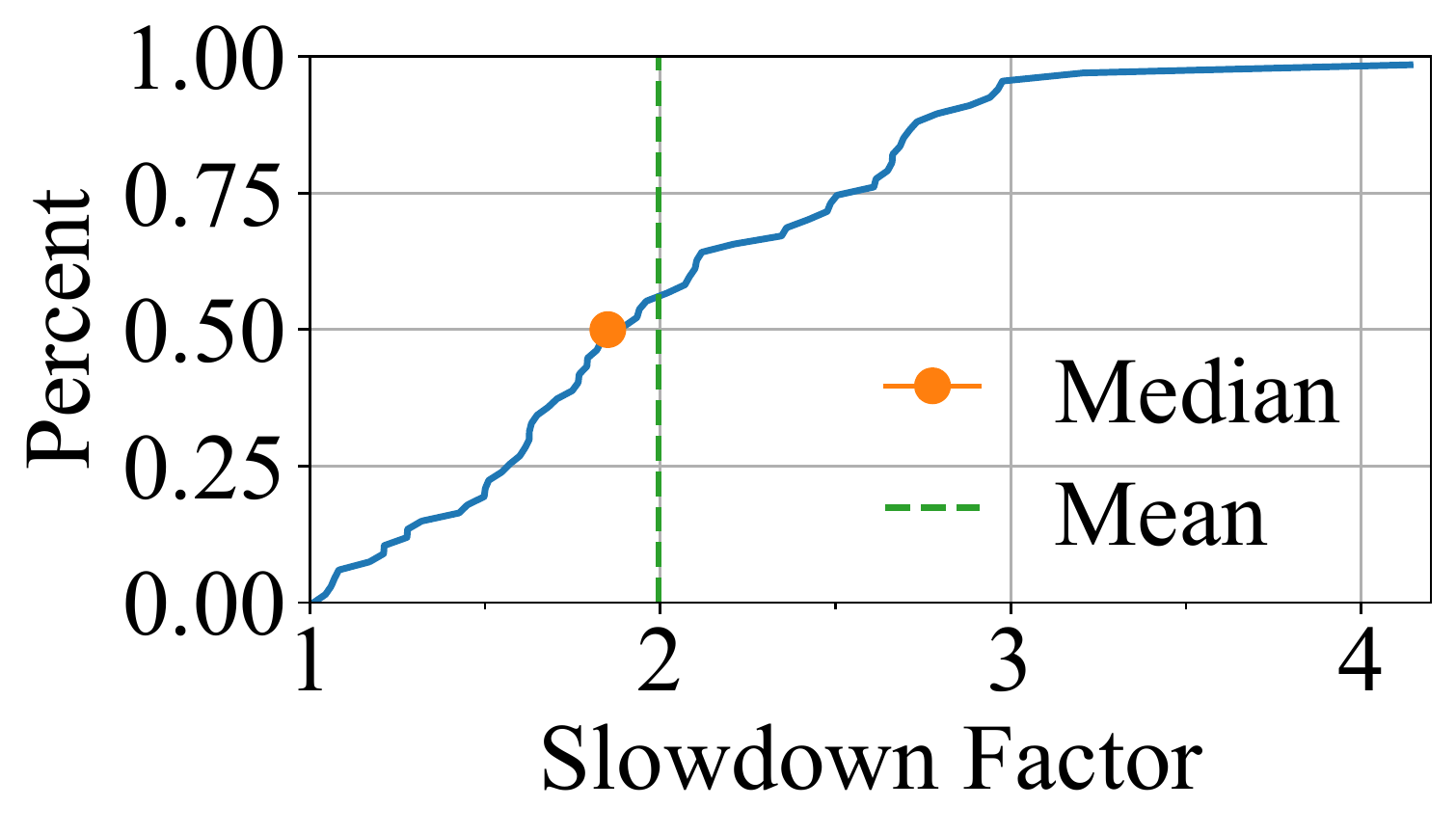}
        \vspace{-3mm}
        \caption{CDF of how much \op is prolonged when it overlaps with \ar operation. 
        We mark the median and average slowdown factors.}
        \label{fig:alltoall_cdf}
    \end{minipage}% 
    \hfill
    \begin{minipage}{0.23\textwidth}
        \centering
        \includegraphics[width=\linewidth]{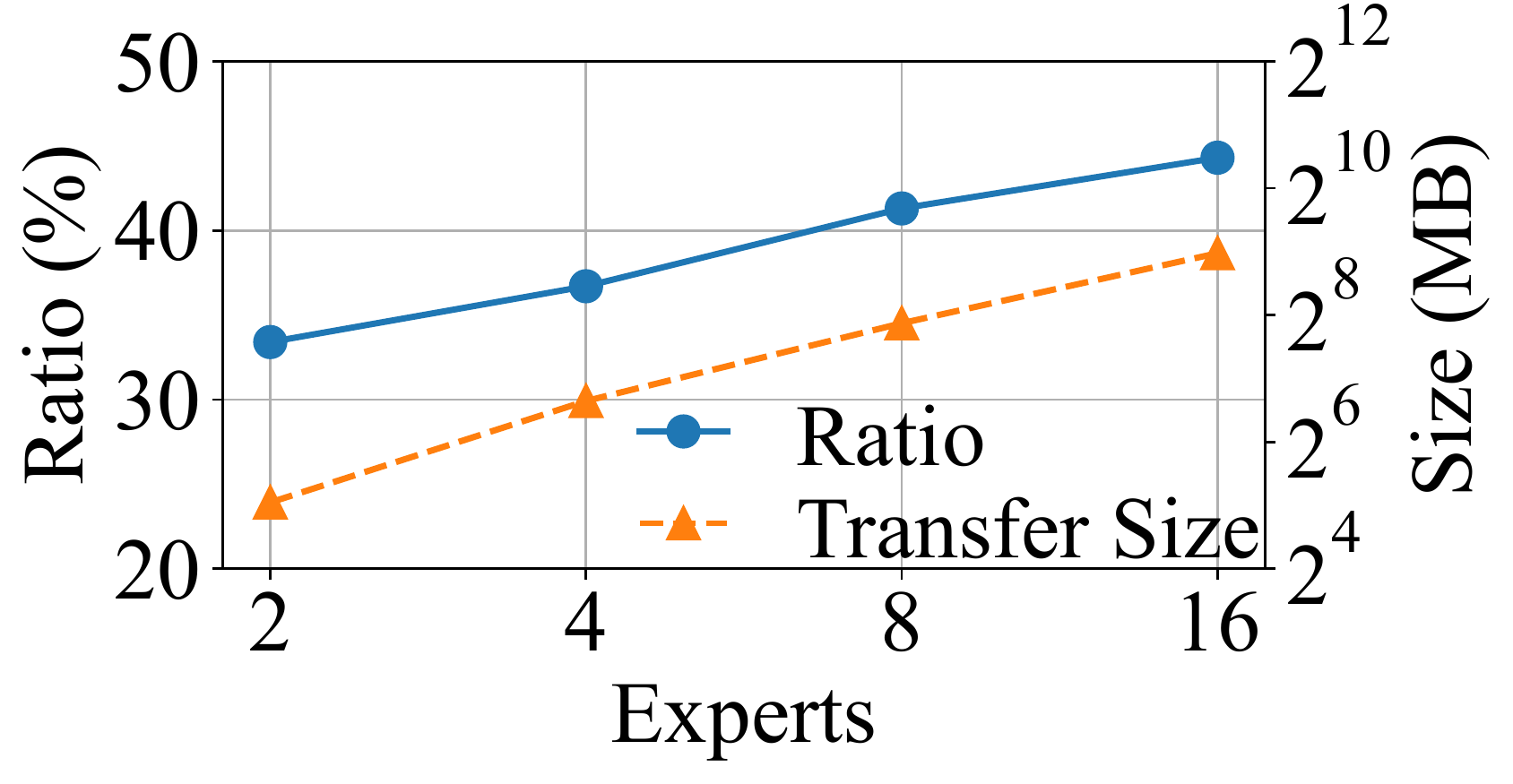}
        \vspace{-8mm}
        \caption{The proportion of \op's completion time over training 
        step time when the number of experts grows. Dashed line plots 
        the data size in one \op operation.}
        \label{fig:alltoall_exp}
    \end{minipage}
    \vspace{-7mm}
\end{figure}

\noindent\textbf{Synchronous \op with large data transfer.} 
The common characteristic shared by MoE training and inference is \op's large data transfer. 
\Op is an irreplaceable synchronous component to handle the data exchange among devices in MoE layer. Each MoE layer has two \op operations to send the tokens to the experts and then restore the position of tokens, as introduced in~\Cref{subsec:background}. The data transfers in the two \op operations have the same size because the expert's FFN architecture ensures that its input data size is the same as the output data size. Figure~\ref{fig:moe_timeline} shows an empirical timeline view of the forward pass of MoE model in our cluster. \Op takes 74.9\% of the end-to-end running time of one MoE layer. 
Expert FFN computation and the combine operation follow when \op operation completes. MoE training and inference suffer from such inefficiency consistently. GPU is mostly idle during this period: We use the PyTorch Profiler~\cite{torchprofiler} to profile the GPU activities for 20 steps in each experiment in Table~\ref{table:alltoall_to_step}, and find that the average GPU SM efficiency during \op is 3.7\%. Besides, the data transfer size grows linearly with the number of experts. Figure~\ref{fig:alltoall_exp} presents the empirical evidence of \op's transfer size as the number of experts grows from 2 to 16 (128). With the increasing number of experts, the time taken by \op grows from 33.4\% to 44.5\% of the step time. 

\begin{figure}[t]
    \centering
    \includegraphics[width=0.85\linewidth]{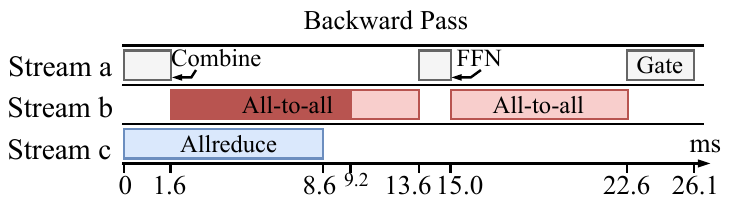}
    \vspace{-2mm}
    \caption{Timeline of backward propagating an MoE layer under hybrid 
    parallelism. The first \op is prolonged by the \ar operation in 
    Stream b. The shadowed part is its original completion time. }
    \label{fig:distributed_moe_timeline}
    \vspace{-2mm}
\end{figure}

% \noindent\textbf{Prolonged \op in training.} 
\noindent\textbf{Problem in training: Prolonged \op with \ar.} 
The unique challenge in MoE training is that applying the hybrid parallelism creates a particularly severe impact to \op in backward pass. Non-MoE layers in \textit{data parallelism} need \textit{\ar} to aggregate the gradients, while \textit{expert parallelism} requires \op to exchange tokens to compute expert gradients. Since the two operations control their own process groups independently, two dedicated CUDA streams are launched concurrently. This is demonstrated in Figure~\ref{fig:distributed_moe_timeline} with the timeline of backward pass in a sample run of MoE training. As the two operations overlap, they contend for the network bandwidth and their completion times are severely prolonged. To make matters worse, we find that the slowdown factor varies significantly. {We collect the completion times of 1,500 \op operations} in backward pass on our testbed and plot the CDF of the slowdown factor they endure with \ar in Figure~\ref{fig:alltoall_cdf}. Observe that the median slowdown is over 1.83x and the worst is 4.14x. 

\begin{figure}[t]
    \centering
    \begin{subfigure}{0.4\textwidth}
        \centering
        \includegraphics[width=0.5\linewidth]{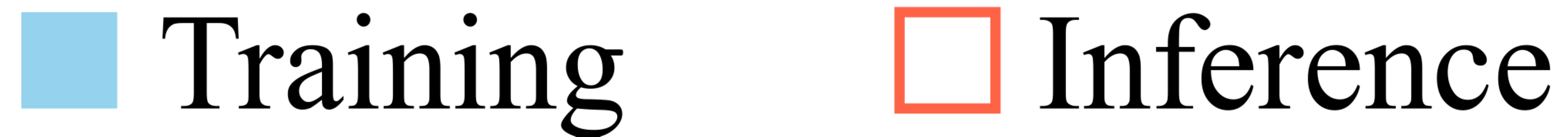}
    \end{subfigure}
    \begin{subfigure}{0.17\textwidth}
        \centering
        \includegraphics[width=\linewidth]{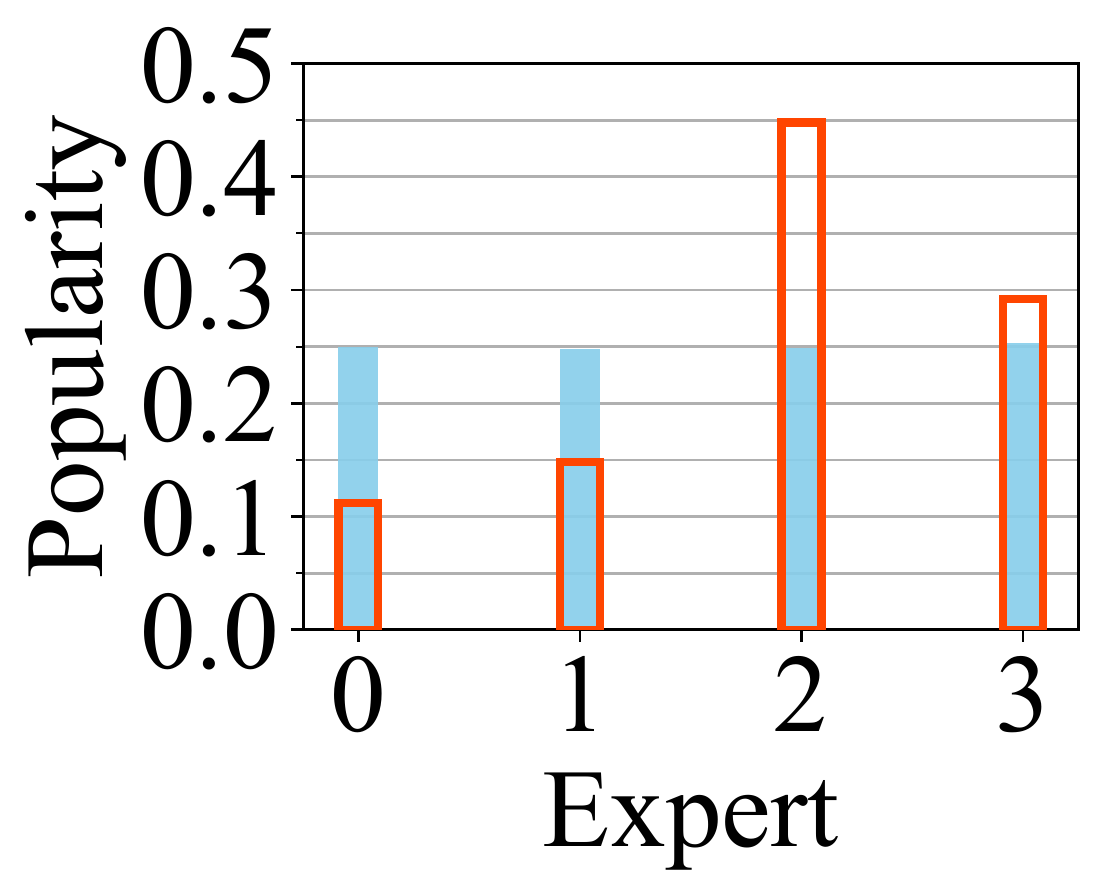}
        \caption{4-expert MoE.}
        \label{fig:exp_popularity_4}
    \end{subfigure}% 
    \hfill
    \begin{subfigure}{0.29\textwidth}
        \centering
        \includegraphics[width=0.9\linewidth]{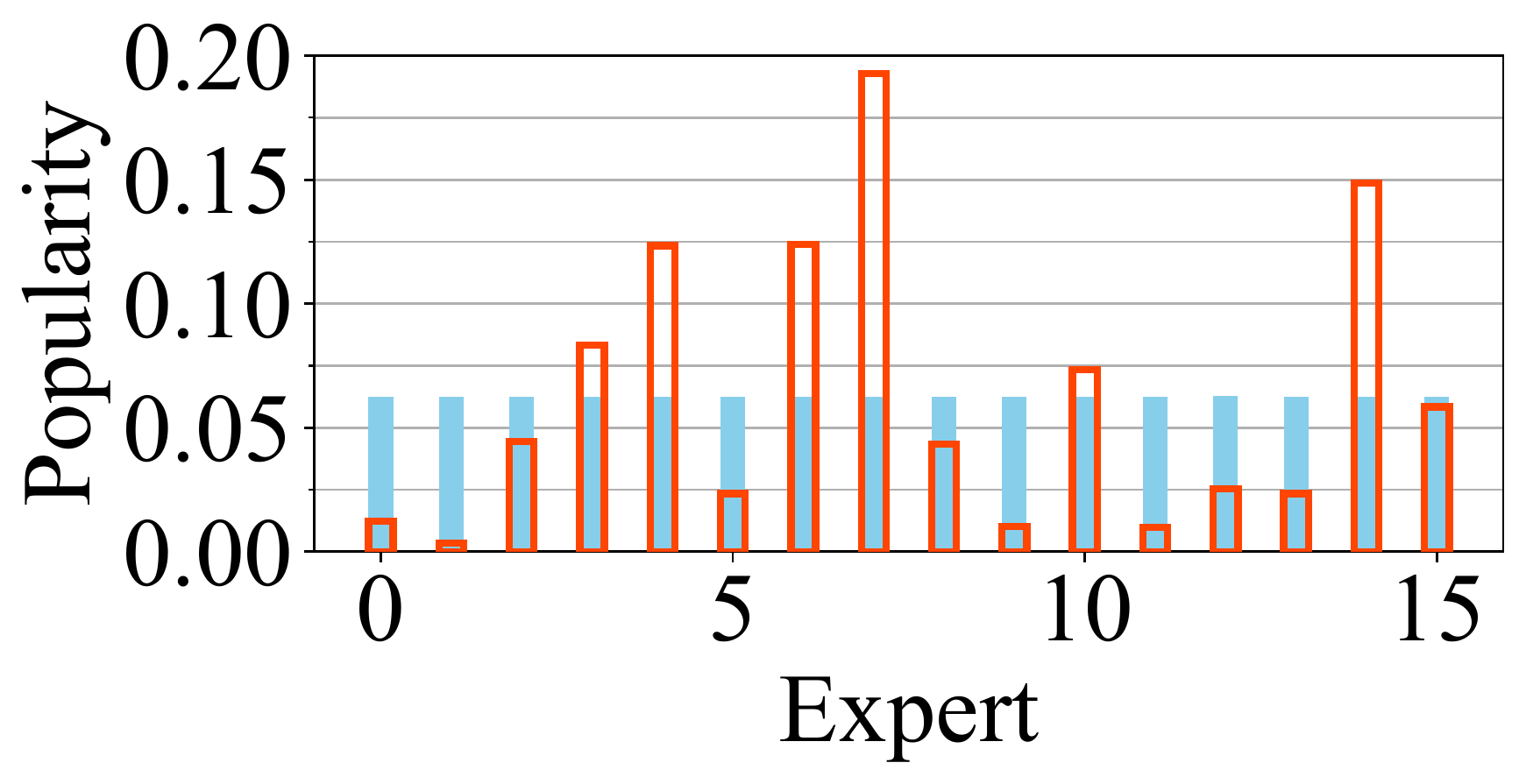}
        \caption{16-expert MoE.}
        \label{fig:exp_popularity_16}
    \end{subfigure}
    \vspace{-2mm}
    \caption{Sampled expert popularity. The distribution is computed as the ratio between the number of tokens received by the expert and total number of tokens in one batch. We use the Enwik8 test set~\cite{wik8} for evaluation.}
    \label{fig:exp_popularity}
    \vspace{-5mm}
\end{figure}

\noindent\textbf{Problem in inference: Skewed expert popularity.} 
The main cause of \op being the bottleneck in MoE inference is the skewed expert popularity. The token-to-expert distribution in inference is purely workload-driven, and we empirically find that the expert popularity is highly {skewed} in sharp contrast to training. We sample the expert popularity of the same MoE model in training and inference in Figure~\ref{fig:exp_popularity}. In training, the distribution is nearly the same across all experts after hundreds of steps due to the use of load balancing loss. In inference, however, the most popular expert receives 4.02x and 5.56x tokens of the least popular ones in 4-expert and 16-expert inference tasks. With the same network and computation capacity, devices hosting popular experts take much longer to perform expert computation. In this experiment, the maximum idle time of the least popular expert is 29.4\% of the inference time of that batch. Thus, within one batch, tokens to the less occupied experts have to wait for others to complete on the more popular experts, degrading the \op performance significantly. Further, under uniform expert-device allocation, devices hosting popular experts have more tokens using their links for \op, while the links of other devices are underutilized.

%% file: design.tex
%!TEX root = main.tex
\section{Design Overview}
\label{sec:design}
\sys is designed to accelerate \op in distributed MoE. 
It attacks both the bandwidth contention with \ar in training, as well as the straggler with unbalanced \op bandwidth in inference. \revise{We focus specifically on MoE implementations that leverage both data and expert parallelism.}

\noindent\textbf{MoE training.} We aim to improve the \textit{bandwidth} of \op in order to reduce the blocking period of the computation operations. Our key idea here is to \textit{prioritize \op} so it does not fair-share bandwidth with concurrent \ar (\Cref{sec:comm_schedule}). This is achieved using tensor-partitioning. We partition \op and \ar tensors into small chunks, each of which then forms a \textit{micro-op}. \sys schedules an \ar micro-op only when there is no \op waiting or ongoing so that \op is guaranteed the full network bandwidth during its lifetime. Without prior information, tensor-partitioning and micro-ops can ensure that in most cases \op can launch immediately and \ar is not deferred excessively. 
 
\noindent\textbf{MoE inference.} We propose to dynamically adjust the device allocation for experts based on the \textit{expert popularity}, so that is not \op is not delayed by the trailing tokens, and its bandwidth utilization across links is balanced (\Cref{sec:inference_schedule}). 
We exploit the expert selection pattern across adjacent layers to estimate the expert popularity.
Based upon the estimation, \sys performs scheduling at each layer to allocates proportionally more devices for popular experts and pack unpopular ones to fewer devices, and coordinate \op correspondingly.

%% file: training.tex
%!TEX root = main.tex
\section{Prioritizing All-to-All Training}
\label{sec:comm_schedule}

We have shown that \op is slowed down significantly if it overlaps with \ar in the backward pass in MoE training. 
\sys partitions the communication operations into small micro-ops and schedule them strategically in order to prioritize \op without impeding \ar and the computation process. 
We introduce the design challenges in~\Cref{subsec:schedule_idea_challenge}.
In~\Cref{subsec:schedule_design}, we present \sys's communication scheduler 
that uses tensor partitioning and pipelining to improve the training 
efficiency.

\subsection{Design Challenge}
\label{subsec:schedule_idea_challenge}

Intuitively, \sys can prioritize \op and avoid concurrent execution with 
\ar with strict priority scheduling. 
\Op is always dispatched first if both are present in the queue, and subsequent operations have to wait until the running one finish to make sure \ar does not share the bandwidth.

\begin{figure}[t]
    \begin{subfigure}{\linewidth}
        \centering
        \includegraphics[width=\linewidth]{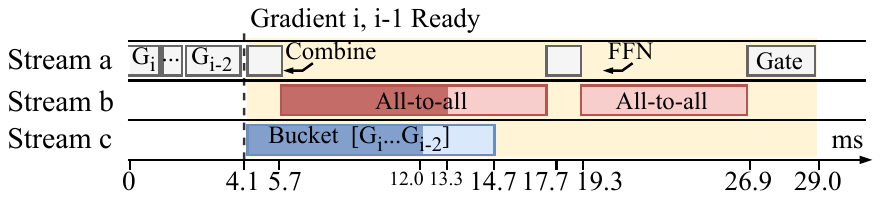}
        \caption{Baseline. Shadowed \op and \ar are their completion times without concurrent operations. Computing the entire MoE layer's gradients ends at 29.0ms.}
        \label{fig:schedule_baseline}
    \end{subfigure}
    \begin{subfigure}{\linewidth}
        \centering
        \includegraphics[width=\linewidth]{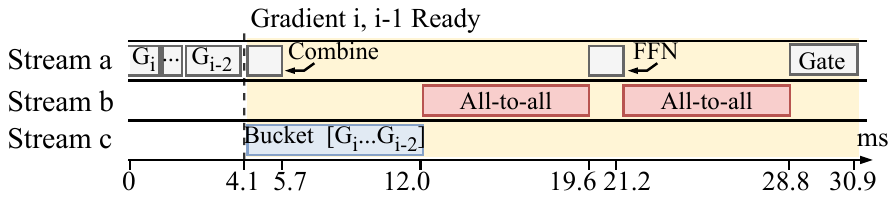}
        \caption{Naively prioritizing \op without concurrent transmission can lead to worse results; computing the MoE layer's gradients ends at 30.9ms. The completion time is profiled. Theoretically, the completion time should be the same as Figure~\ref{fig:schedule_baseline}.}
        \label{fig:schedule_naive}
    \end{subfigure}
    \begin{subfigure}{\linewidth}
        \centering
        \includegraphics[width=\linewidth]{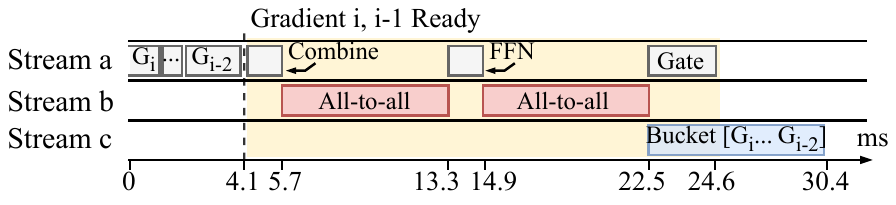}
        \caption{Deferring \ar to after the second \op leads to better training efficiency; computing the MoE layer's gradients ends at 24.6ms.} %
        \label{fig:schedule_optimal}
    \end{subfigure}
    \begin{subfigure}{\linewidth}
        \centering
        \includegraphics[width=\linewidth]{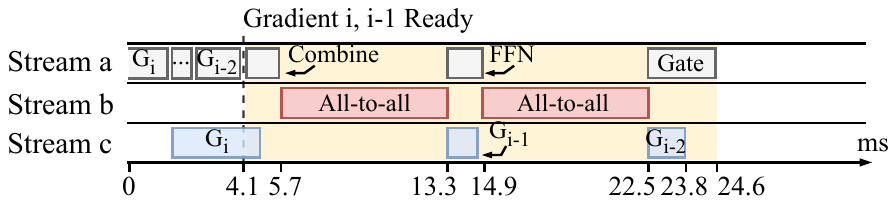}
        \caption{Scheduling results if the arrival time and running time of communication 
        operations are known a priori. The \ar completes much faster than (c).}
        \label{fig:schedule_optimal_short}
    \end{subfigure}
    \vspace{-3mm}
    \caption{
        Backward pass of MoE training. The yellow background is the period of computing the gradients of the MoE layer. Stream a is responsible for the computation process and streams b and c are for communication. 
        This timeline is extracted from a real run of the 419M-parameter benchmark model in Table~\ref{table:alltoall_to_step}.}
\vspace{-6mm}
\end{figure}

It turns out that simply prioritizing \op is not as efficient as one may expect. 
For work-conservation, when an \ar arrives first, it should be launched immediately. The problem is when an \op arrives later, though ideally one would preempt the \ar due to priority scheduling, this is not possible in current multi-GPU communication libraries such as NCCL~\cite{nccl}.  
The communication primitives are highly optimized and upon being called, their complete transmission strategies are settled and pushed to the CUDA streams. There is no control knob inside each primitive to adjust how it shares resources (e.g. CUDA cores, network bandwidth) with others.
Thus, as the example in Figure~\ref{fig:schedule_naive} shows (based on testbed experiments), naively prioritizing \op actually leads to a longer completion time for the first \op and training step time compared to the baseline in Figure~\ref{fig:schedule_baseline}.

\begin{figure}[t]
    \begin{subfigure}{\linewidth}
        \centering
        \includegraphics[width=\linewidth]{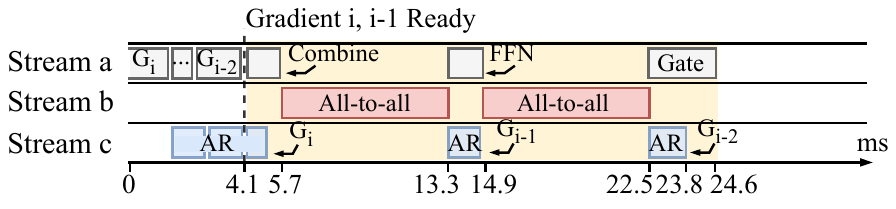}
        \caption{Prioritize \op and partition \ar tensors. 
        Instead of bucketing gradients, we partition gradient $i$ into three 
        chunks when it is computed.}
        \label{fig:schedule_partition}
    \end{subfigure}
    \begin{subfigure}{\linewidth}
        \centering
        \includegraphics[width=\linewidth]{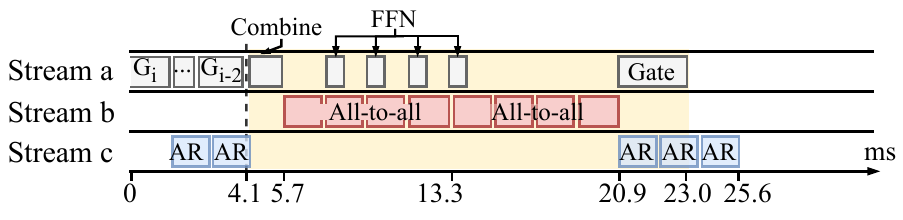}
        \caption{Tensor partitioning for \op and pipeline the FFN computation. }
        \label{fig:schedule_pipeline}
    \end{subfigure}
    \caption{We show the scheduling results from Figure~\ref{fig:schedule_baseline} with tensor partitioning. \Op and \ar micro-ops are of the same size.} 
    \vspace{-5mm}
\end{figure}

A potential solution is to obtain the arrival time and running time of the upcoming \op and \ar, and orchestrate them accordingly to maximize the efficiency. 
Assuming we know that the \ar for gradient $i$ can complete before \op and the completion 
time of gradient $i-1$'s \ar is shorter than FFN computation. 
Then we can schedule gradient $i-1$'s \ar to the gap between the two \op operations at 13.3ms as depicted in Figure~\ref{fig:schedule_optimal_short}. 
Obtaining the precise knowledge of arrival and running times is, however, a daunting task. 
ML frameworks such as PyTorch fuse gradients into buckets based on a user-defined bucket size to optimize \ar efficiency. 
Yet in large Transformer-based models, gradient sizes are also large; since bucketing is done on the gradient boundary, the actual bucket size for \ar varies wildly~\cite{torchddp}. 
Moreover, the implementation details of \ar make it difficult to 
acquire a reliable running time estimate as prior work has found out~\cite{MLSYS2019_9b861925}.

The other design choice is to blindly defer \ar until an even number of
\op finish as there should be a larger gap between the backward pass of two
MoE layers relative to FFN's backward computation.
Figure~\ref{fig:schedule_optimal} shows the best scheduling result based on the
baseline in Figure~\ref{fig:schedule_baseline}. In this case, \ar can be launched when the second \op finishes and completes before the first \op of the next MoE layer (not shown in the figure). Yet, in other (worse) cases, \ar may still block the \op of the upcoming MoE layer if it takes relatively longer.
In the extreme case, no \ar can be launched until all four  
\op operations of the current step finish. 
Since devices have to wait for \ar before moving onto the optimization phase, this incurs more delay and is undesirable for wait-free backward pass~\cite{zhang2017poseidon}.

\subsection{Tensor Partitioning and Micro-Ops}
\label{subsec:schedule_design}
To resolve the above challenges, we propose tensor partitioning that breaks down a communication operation into micro-ops, which can be easily prioritized with high efficiency.

\noindent\textbf{Tensor partitioning.} 
Unlike tensor bucketing which fuses multiple gradients for an \ar, \sys partitions each gradient tensor into equal-sized small chunks and executes individual \ar \textit{micro-ops} independently. 
This brings two advantages. 
First, it resolves the varying bucket size problem for \ar since each micro-op is uniform in size now. %
Second, micro-ops naturally make better use of bandwidth \cite{OPRV13} without causing too much delay to \ar under priority scheduling. 
Consider the same setup from Figure~\ref{fig:schedule_baseline}, in Figure~\ref{fig:schedule_partition} we partition gradients into five chunks. 
Before the first \op arrives, \sys launches three \ar micro-ops; after the first \op ends, it starts another micro-op to opportunistically make use of the expert computation time. 
Compared to the scheduling result without micro-ops in Figure~\ref{fig:schedule_optimal}, \ar for gradient $i-2$ now completes 6.6ms or 21.7\% faster without prolonging \op. 
Tensor partitioning does incur overhead due to the partition and concatenation operations before and after an \ar, but it is mild: the overall overhead in Figure~\ref{fig:schedule_partition}'s case is 764us. 
\Cref{subsec:scheduler_perf} has more details of the overhead analysis.

\noindent\textbf{Pipelining micro-ops.} 
Intuitively, we can also partition \op which provides an opportunity to pipeline the expert FFN and further reduce the time that computation is blocked. 
Specifically, we can pipeline the expert computation and \op micro-ops (Figure~\ref{fig:schedule_pipeline}). 
Since the FFN computation is in token granularity, the expert can start computing with a subset of the tokens after one \op micro-op. With pipelining, 
we can eliminate the FFN time which is 1.6ms in this example. 

\noindent\textbf{Expert packing.} Ideally, the FFN and \op micro-ops should take a similar time so that both compute capacity and network bandwidth are fully utilized without any bubbles in the pipeline. However, we notice that a single FFN micro-op takes much less time than its corresponding \op micro-op (Figure~\ref{fig:schedule_pipeline}). In \sys, we consider packing multiple experts on each device whenever possible to maximize the pipelining efficiency. \sys adopts the following approach: starting with one expert per device, it iteratively increases the number of experts per device in powers of two, until the FFN computation exceeds that of the \op micro-op. \revise{In case of GPU memory shortage, we adopt DRAM-offloading~\cite{ren2021zero} to transfer expert parameters that are not currently in use to host memory.}

%% file: inference.tex
%!TEX root = main.tex
\section{Scheduling Resources in Inference}
\label{sec:inference_schedule}
Recall in~\Cref{subsec:bottleneck}, we have shown empirically that skewed
expert popularity leads to unbalanced processing times across tokens of the
same batch in MoE inference, which delays \op and causes imbalanced bandwidth for it severely. The root cause lies in the data granularity mismatch between the expert and the
attention layers in the model: an expert processes individual
tokens, but the attention layer processes an entire sequence as a whole.
Our design question is thus: How can we ensure that each token within the
same batch experiences the same end-to-end completion time no matter its
expert selection result? 
We will first discuss the challenge of
achieving this through dynamic resource scheduling (\Cref{subsec:inference_challenge}), and then present our design that exploits the unique token-level expert selection pattern to address the challenge in~\Cref{subsec:inference_design}.  

\subsection{Design Challenge}
\label{subsec:inference_challenge}

To cope with skewed expert popularity, intuitively one must accordingly adjust the resource allocation for experts. 
This adjustment also needs to be done for each input sequence as the expert popularity distribution varies across sequences.
An immediate question is: how can we know the expert popularity distribution, before the input is processed by the gating network?

This question is challenging for two reasons.
First, even for a given batch of input, expert popularity varies across MoE layers of the model. 
We collect the expert popularity of different MoE layers for 1000 batches of input requests. 
Table~\ref{table:top_4_exp_pop} shows the top-4 popular experts of two 12-expert inference tasks: text generation and translation. 
Observe that each MoE layer of the same task (model) has completely different popular experts. 
This also suggests that dynamic resource scheduling has to be done before each MoE layer in order to be effective.
Moreover, scheduling resources according to the actual expert selection results, as some might be thinking, incurs delay in collecting information, making scheduling decisions, and coordinating the all-to-all amongst all experts with respect to the new expert-device mapping, all of which are blocking operations and are performed at each layer. This is far from optimal (as will be shown in~\Cref{subsubsec:inference_sched_perf}).
Thus, we need to know as much as possible the expert popularity \textit{before} the gating network selects experts in each layer, so these overheads can be largely overlapped with MoE computation.

\subsection{Popularity based Scheduling}
\label{subsec:inference_design}

\sys tackles the design challenge by exploiting the token-level expert selection pattern which we empirically establish now. Building upon this, we design a resource scheduler that replicates popular experts on proportionally more devices in order to better balance the workload.

\begin{figure}[t]
    \begin{minipage}{0.23\textwidth}
        \centering
        \resizebox{\columnwidth}{!}{
        \begin{tabular}{@{}lccccc@{}}
            \toprule
            Model\& Dataset & Layer & \multicolumn{4}{c}{Top-4} \\ \midrule
            \multirow{4}{*}{\begin{tabular}[c]{@{}l@{}l@{}}Transformer-XL\\ \& Enwik8 \\ (Text generation)\end{tabular}} & 3   & 9   & 4    & 5   & 10    \\
            & 4     & 5    & 7    & 8    & 10   \\
            & 8     & 9    & 2    & 3    & 13   \\
            & 12    & 4    & 5    & 15   & 8    \\ \midrule
            \multirow{4}{*}{\begin{tabular}[c]{@{}l@{}l@{}}BERT-Large\\ \& WMT En-De \\ (Translation)\end{tabular}} & 6   & 7   & 6    & 10   & 1   \\
            & 8    & 10   & 6    & 2    & 15   \\
            & 10   & 9    & 4    & 11   & 8   \\
            & 12   & 1    & 8    & 10   & 14    \\ \bottomrule
        \end{tabular}}
        \captionof{table}{Top-4 popular experts in sampled MoE layer of two MoE models.}
        \label{table:top_4_exp_pop}
    \end{minipage}%
    \hfill
    \begin{minipage}{0.23\textwidth}
        \centering
        \includegraphics[width=\linewidth]{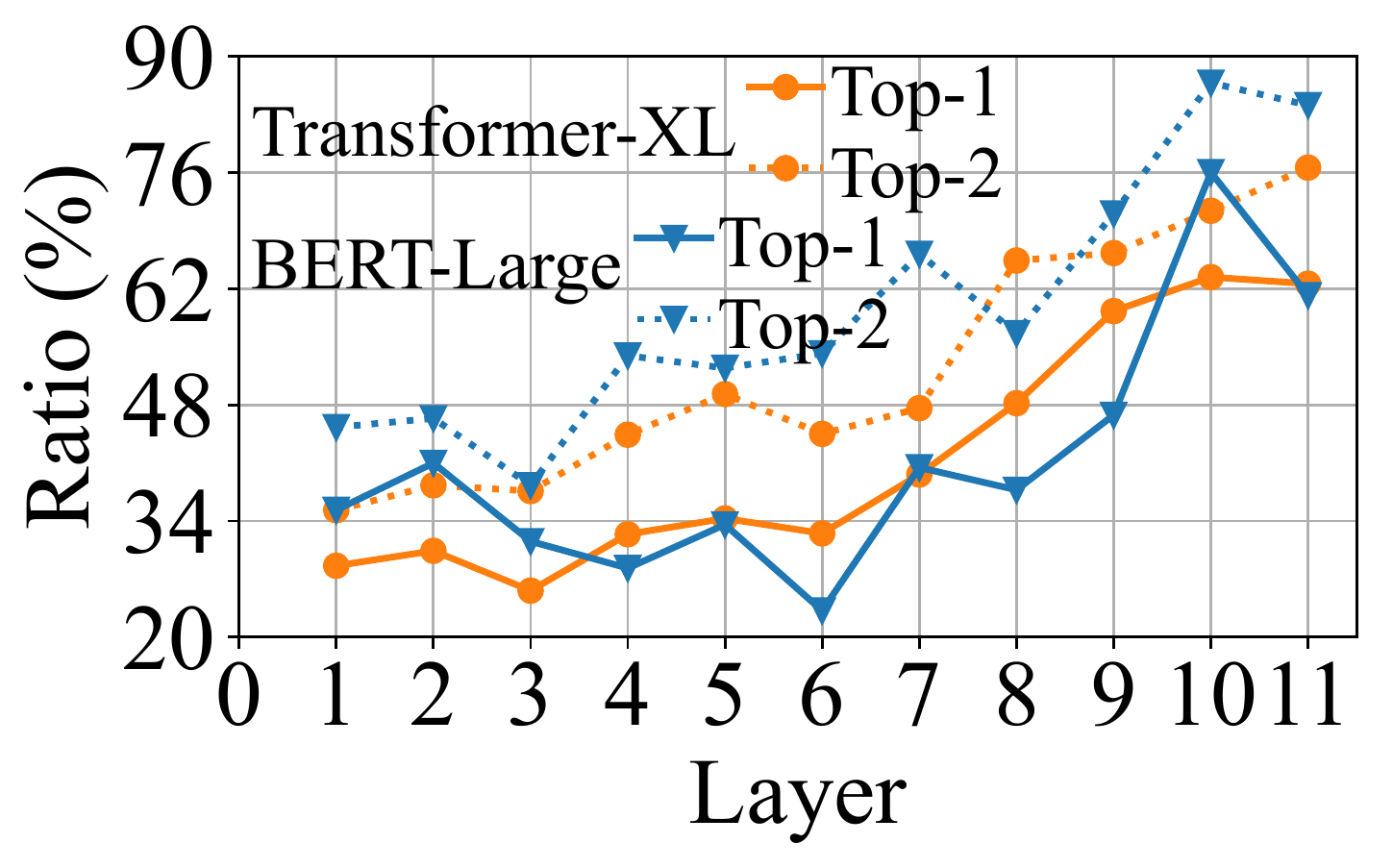}
        \vspace{-7mm}
        \caption{Ratio of tokens that select one of the top-$k$ experts in layer $i+1$ given that they have selected the same expert in layer $i$. }
        \label{fig:popularity_pattern}  
    \end{minipage}
    \vspace{-7mm}
\end{figure}

\noindent\textbf{Pattern in expert selection.} 
Experts in MoE models are trained to specialize in different types of input. We find that a token's expert selection demonstrates a pattern across the MoE layers. Tokens that have selected the same expert in layer $i$ tend to select the same expert again in layer $i+1$. 
We trace the expert selection of sampled tokens.
For each group of tokens that have selected the same expert in layer $i$, we calculate the ratio of them that in the next layer also select one of the same top-$k$ experts ranked locally among the same group. 
Figure~\ref{fig:popularity_pattern} plots this ratio averaged over token groups in two 12-layer MoE models. 
We see 41.94\% tokens exhibit this pattern when $k$ is 1 and 54.59\% when $k$ is 2, and deeper layers see more tokens with this pattern. 

This observation makes intuition sense. The gating network has a simple architecture, and their routing or expert selection decision is made (largely) based on relatively simple features, such as the parts of speech of a word (noun, verb, etc.), and the meaning of the word (number, time, etc.)~\cite{pmlr-v139-lewis21a}. These features are fixed for each token. Meanwhile, experts focus on the local syntax information of each token rather than the cross-dependency within a sequence. For all these reasons, similar tokens naturally tend to be processed by the same or similar experts in each layer.
% \hx{intuition? why this makes sense}

\noindent\textbf{Estimating expert popularity.} 
Though this pattern may not be sufficient to predict a particular token's expert selection accurately, it provides enough clues for us to estimate the overall expert popularity for a given batch. 
Specifically, \sys's estimation approach works as follows. 
In the profiling stage, we collect the expert selection results of all tokens when the load balancing loss is minimized and becomes stable.
We then group tokens that select the same experts from layer $i-l$ to layer $i$, which represent a unique sample path of experts used. 
For each sample path $j$, we compute the expert popularity distribution $\Psi_j^{i+1}$ for layer $i+1$. Here $l$ is the path length to control the accuracy-cost tradeoff in profiling: a larger path length leads to more accurate estimation for layer $i+1$ at the expense of higher data collection and computation costs.  

Then based on the profiled distributions $\{\Psi\}$, \sys can estimate the next layer's expert selection distribution for each sample path of experts traversed by a token in inference (starting from the $l$-th layer of the model).
In each layer $i$, for a sample path $j$, we pick the top-$k$ expert(s) of the subsequent layer from $\Psi_j^{i+1}$ and use their probabilities $\{P_j^{i+1}(e)\}$ to represent expert popularity for resource scheduling, where $e$ denotes an expert.
The reason why we only consider top-$k$ experts is that they demand the most resources, and the remaining experts have low popularity (Figure~\ref{fig:popularity_pattern}).
Note that this estimation happens before any MoE layer computation takes place.

\noindent\textbf{Two-phase scheduling.} 
During inference, \sys dynamically conducts layer-wise resource scheduling in two phases. 

The first phase happens right after the expert popularity estimation at each MoE layer, when \sys relies on the estimation to replicate popular experts on more devices and pack unpopular ones onto fewer devices.
Specifically, the total number of devices for expert $e$ is determined by: 
\begin{equation}
    n_e = {N\times \sum_{t=1}^{N_t} P^{i+1}_{j(t)}(e)/N_t},
    \label{eqn:device_exp}
\end{equation}
That is, for the current batch of input with $N_t$ tokens, using estimation from each token $t$'s sample path $j(t)$ up to layer $i$, the overall popularity of expert $e$ is estimated as $\sum_{t=1}^{N_t} P^{i+1}_{j(t)}(e)/N_t$ for layer $i+1$ accounting for all tokens.
This requires the same proportion of devices assuming the expert parallelism degree is 1 (i.e. the number of devices equals the number of experts). 
For experts with the estimation $n_e$, we adopt the first-fit-decreasing heuristic to pack them into the empty devices so the total devices used are minimized. 
It is possible that some experts, being extremely unpopular (for this batch), are not amongst the top-$k$ list of any tokens and thus do not have their $n_e$ estimation.
They are assigned evenly to the remaining free devices if any; otherwise are randomly assigned to a device.  

In phase two, \sys fine-tunes the estimation-based scheduling decision after the gating network selects the actual experts. 
It checks if the selection result deviates significantly from the estimation, by comparing the overall top-$2k$ experts. If the two lists are identical, no fine-tuning is needed and inference continues. 
Otherwise, the scheduler re-computes the resource allocation with the actual expert popularity now available following the same logic in phase 1.
The fine-tuning phase does incur delay to collect the gating outputs and check against the estimation, which is necessary to deal with inaccurate estimation that turns out to be much more detrimental to performance, if left unchecked (\Cref{subsec:inference_perf}). 

%% file: implementation.tex
%!TEX root = main.tex
\section{Implementation}
\label{sec:implementation}
We implement \sys on DeepSpeed MoE and PyTorch using C++ and Python. 
PyTorch 1.10, CUDA 11, and NCCL 2.10 are used.
We modify PyTorch's implementation of distributed training to support \sys in DeepSpeed.  
The implementation has $\sim$7500 LoC. 

\subsection{Training}
\label{subsec:training_impl}
\sys's communication scheduler for training is deployed on all devices and runs a single thread. 
Since the communication scheduling is purely local in scope, no coordination is needed across the scheduler instances on different devices.

\noindent\textbf{Communication scheduler.} 
Each scheduler instance maintains a priority queue to schedule the micro-ops. 
The micro-op size is passed in as a hyperparameter. \sys uses the built-in APIs \texttt{chunk} and \texttt{cat} in LibTorch to partition the data in the token dimension.
We avoid putting chunks from different gradients into the same micro-op to simplify the subsequent concatenation operation. 
Moreover, the scheduler stops launching \ar micro-ops if the combining computation in backward pass, since this implies \op is imminent. We pipeline \op micro-op in the MoE layer. FFN is ready to start right after each \op micro-op. 

\noindent\textbf{Expert packing coordinator.}
We embed a packing controller in the MoE model and it runs a single thread. Expert packing is dynamically adjusted after 10 training steps. In the forward pass, the controller records the completion times of \op and FFN micro-ops. When FFN micro-ops are shorter than \op, the controller starts to pack experts. First, we initialize the new process groups. Second, the controller inserts a one-time synchronous \op to exchange expert parameters between packed devices that would be invoked at the upcoming iteration. Finally, multi-stream parallel execution is adopted for both forward and backward passes when more than one expert are hosted on a device. 

\subsection{Inference}
\label{subsec:inference_impl}
\noindent\textbf{Resource scheduler.} 
The inference scheduler runs on a dedicated thread on device 0 of the cluster and manages resource scheduling.
Each device saves the weights of all experts in their host DRAM and the collected layer-wise expert popularity distribution using multiple \texttt{unordered\_map}, one for each layer. If GPU memory is in shortage, a device only loads one expert and the profiled distribution of one layer at a time. 

In phase one of scheduling, all relevant communication happens by piggybacking the information on the regular \op to reduce overheads.
For each MoE layer, each device appends the popularity estimation to the first \op for device 0.
The scheduler computes the new expert-device mapping and instructs each device which expert and how many to host via the second \op. 
We also include necessary information to coordinate \op of the next layer, including the list of devices with the same expert, and how many tokens each replica should handle to balance the load. 
Devices then swap in the expert weights for the next layer. 
All these procedures are pipelined with model computation.

In phase two, each device updates the actual expert popularity in a separate NCCL \texttt{send} to the scheduler. 
If no fine-tuning is required, the scheduler broadcasts a resume signal. This only creates a negligible overhead as the transfer size is tiny. 
Otherwise, \sys broadcasts the fine-tuned expert-device mapping. 
The model computation is blocked during phase two until the scheduler's command is received.

\noindent\textbf{\Op coordination.} In inference, \sys uses \op with an unequal split. That is, the transfer size to each device in \op does not need to be the same. Using unequal split \op can save the overhead of initializing multiple process groups. A placeholder data pointer is passed to \op if no tokens are directed to a certain device. 

\noindent\textbf{Expert packing.} 
Expert computation is sequential on devices hosting multiple experts. Each device loads the experts one at a time to perform computation and move on to the next packed expert. \revise{In this manner, \sys avoids placing extra strain on the GPU memory.}
The second \op is launched when the computation for all packed experts is completed. 
We set a maximum number of experts per device to control the overhead of swapping the weights. 

%% file: evaluation.tex
%!TEX root = main.tex
\section{Evaluation}
\label{sec:evaluation}

We present the testbed evaluation results here. 

\subsection{Setup}
\label{subsec:eval_setup}
\noindent\textbf{Testbed setup.} 
Our testbed has four worker nodes. Each node has 4 Ampere A100 GPUs with 40GB memory and is equipped with 100Gbps InfiniBand.  

\noindent\textbf{MoE models.} We convert three common Transformer-based dense language models to MoE ones for training. 
\begin{itemize}[leftmargin=*]
    \item Transformer-XL~\cite{dai2019transformer}: a 24-layer encoder model. 
    \item BERT2GPT2~\cite{wolf-etal-2020-transformers}: a 12-layer encoder-decoder model. 
    \item GPT-2~\cite{radford2019language}: a 12-layer decoder model.
\end{itemize}

\noindent Besides, we consider two inference tasks. 
\begin{itemize}[leftmargin=*]
    \item Transformer-XL~\cite{dai2019transformer}: The inference task is text generation with Enwik8~\cite{wik8} test set. 
    \item BERT-Large~\cite{devlin2018bert}: a 12-layer decoder model. The inference task is translation using WMT En-De~\cite{wmt} test set.
\end{itemize}

All FFN layers in these models are converted to MoE layers. We vary the number
of experts in an MoE layer from 2, 4, 8, to 16. We adopt top-2 gating in
training and top-1 gating in inference following~\cite{fedus2021switch}, i.e. $k=2$ in training and $k=1$ in inference 

\begin{figure*}
    \centering
    \includegraphics[width=0.36\linewidth]{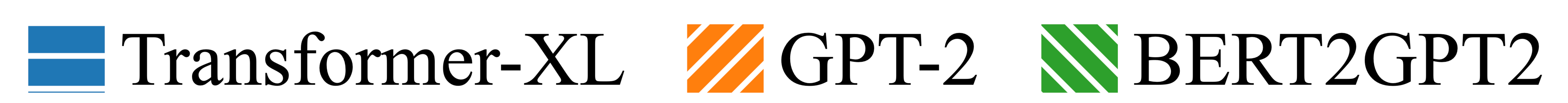}
    \vspace{-5mm}
\end{figure*}
\begin{figure*}
    \begin{minipage}{\textwidth}
    \begin{minipage}{0.24\textwidth}
        \centering
        \includegraphics[width=\linewidth]{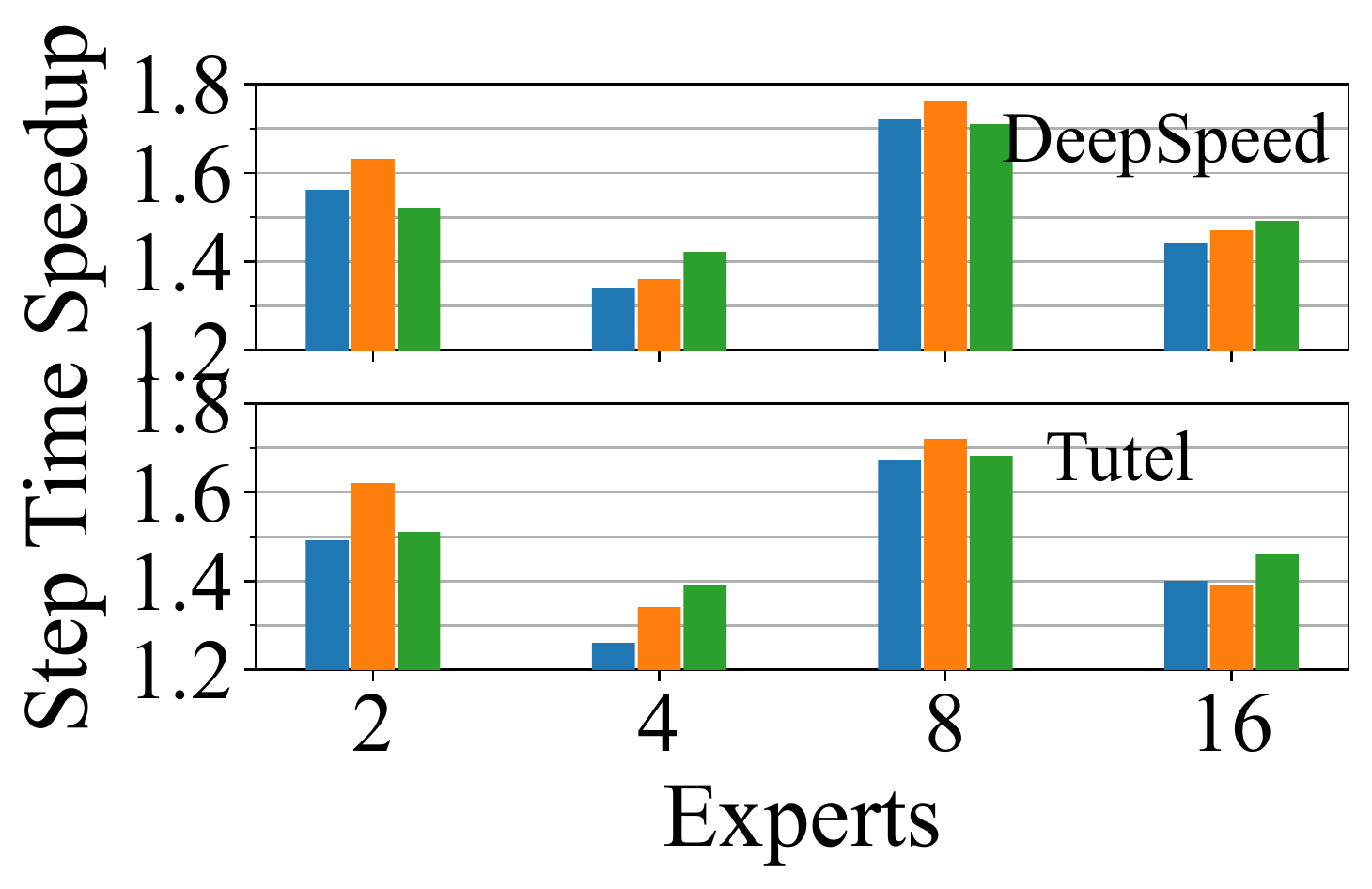}
        \vspace{-8mm}
        \caption{Speedup of training step time against two Baselines.}
        \label{fig:overall_step_speedup}
    \end{minipage}%
    \hfill
    \begin{minipage}{0.24\textwidth}
        \centering
        \includegraphics[width=\linewidth]{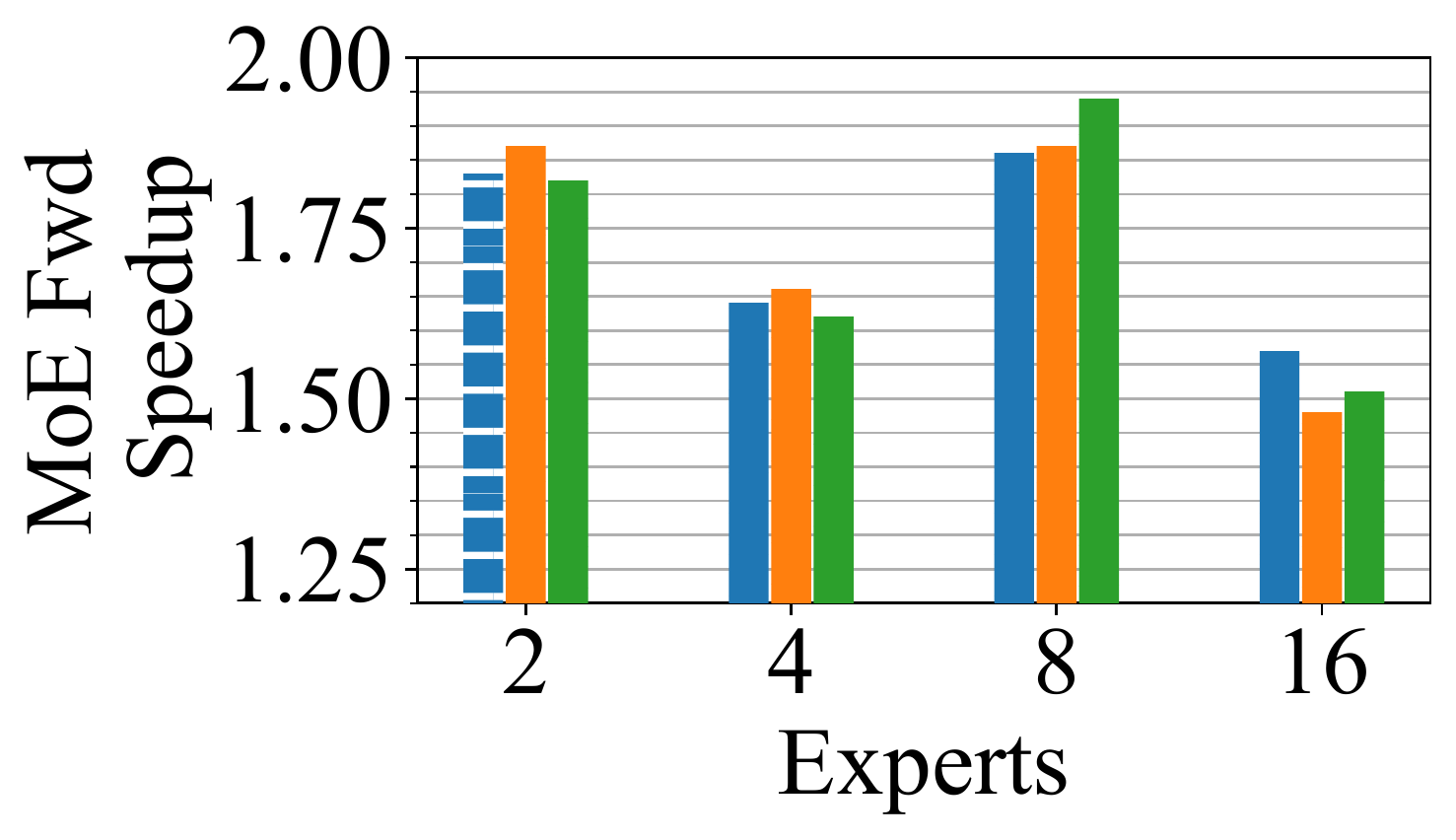}
        \vspace{-8mm}
        \caption{Speedup of MoE layer's forward pass completion time.}
        \label{fig:overall_moefwd_speedup}
    \end{minipage}%
    \hfill
    \begin{minipage}{0.24\textwidth}
        \centering
        \includegraphics[width=\linewidth]{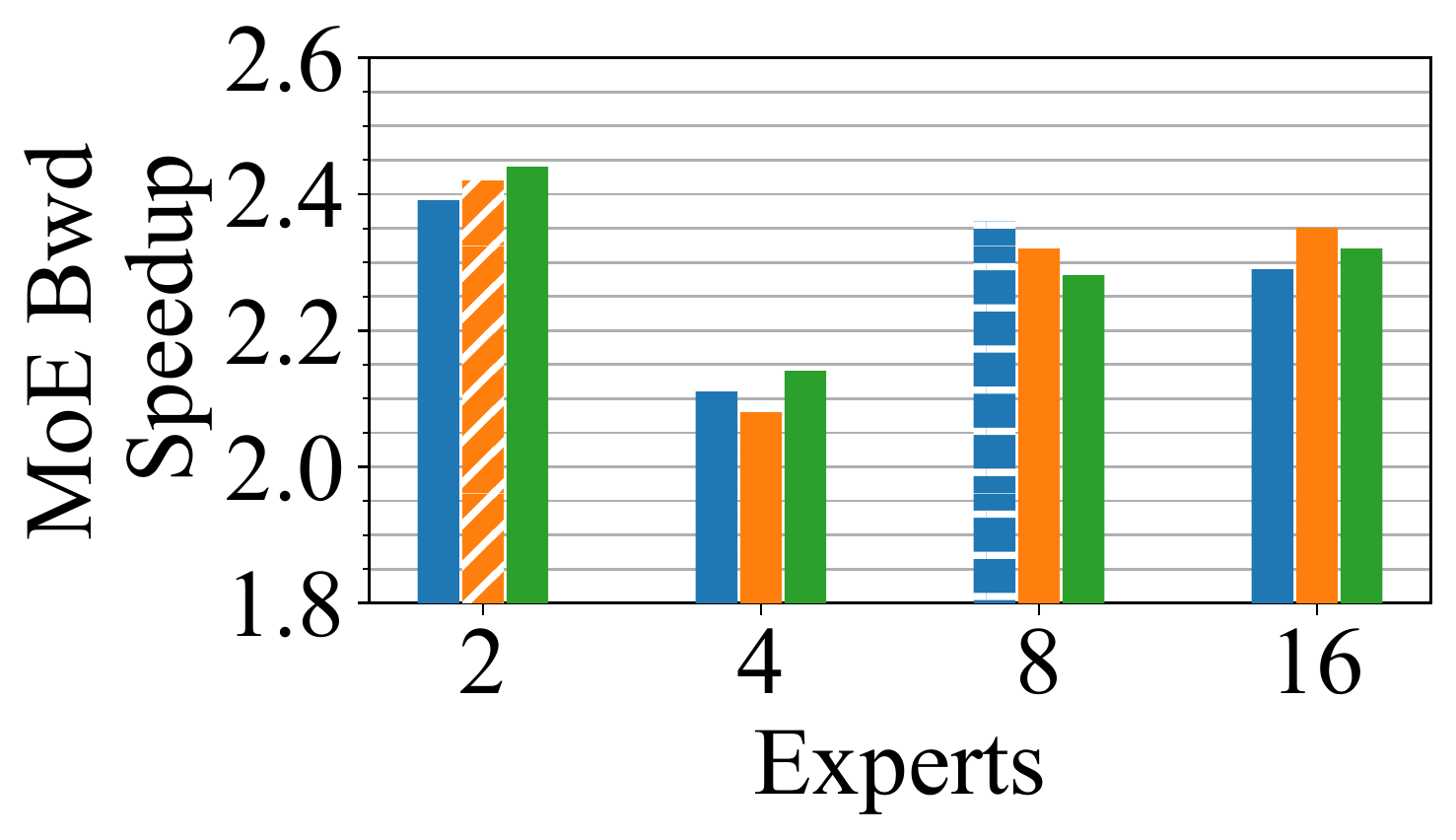}
        \vspace{-8mm}
        \caption{Speedup of MoE layer's backward pass completion time.}
        \label{fig:overall_moebwd_speedup}
    \end{minipage}%
    \hfill
    \begin{minipage}{0.24\textwidth}
        \centering
        \includegraphics[width=\linewidth]{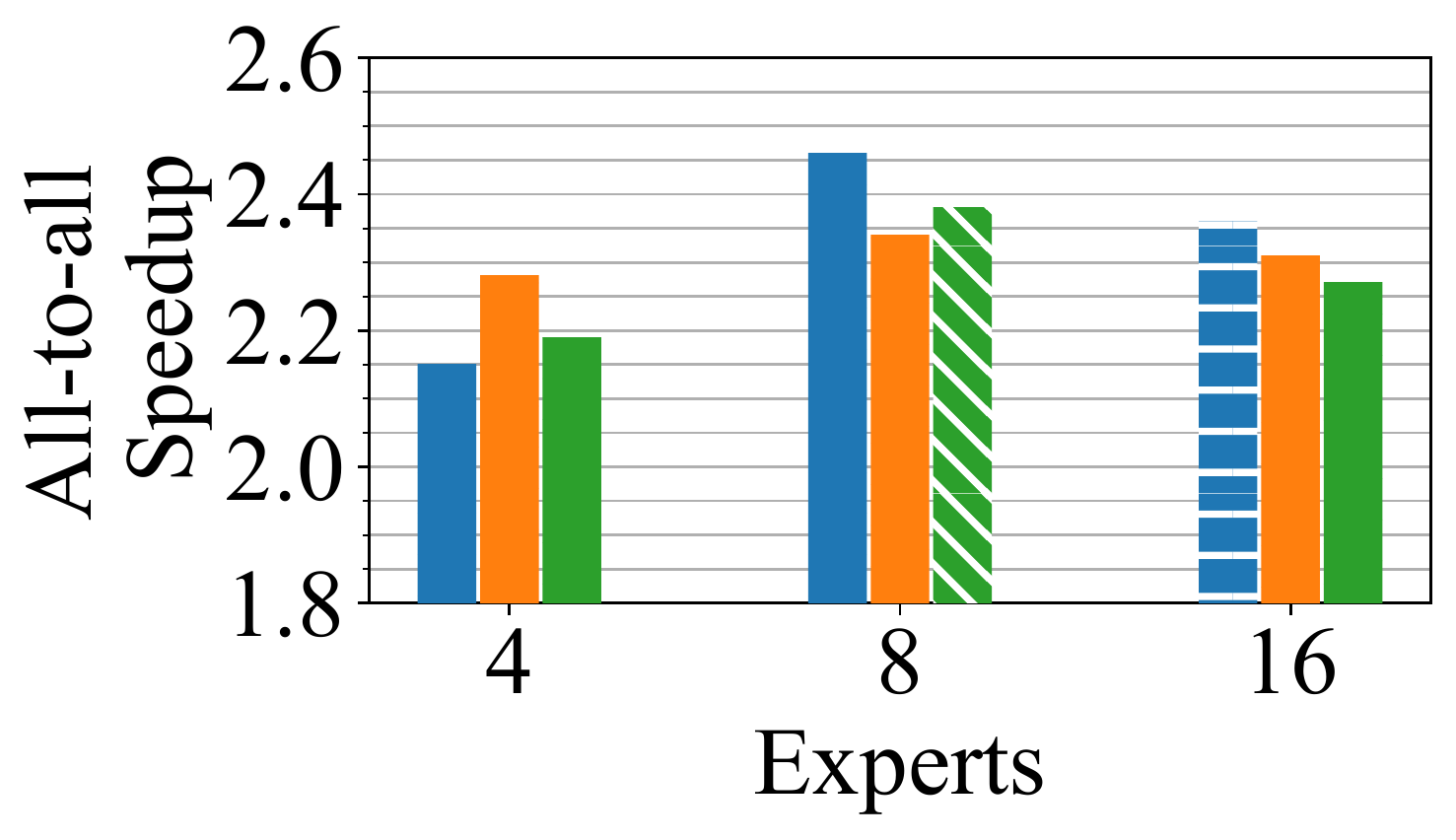}
        \vspace{-8mm}
        \caption{Speedup of \op time in forward and backward pass.}
        \label{fig:overall_a2a_speedup}
    \end{minipage}
\end{minipage}
    \begin{minipage}{\textwidth}
        \centering
    \includegraphics[width=0.68\textwidth]{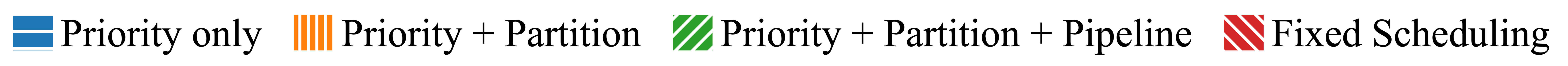}
    \end{minipage}
    \begin{minipage}{0.74\textwidth}
    \begin{subfigure}{0.33\linewidth}
        \centering
        \includegraphics[width=\linewidth]{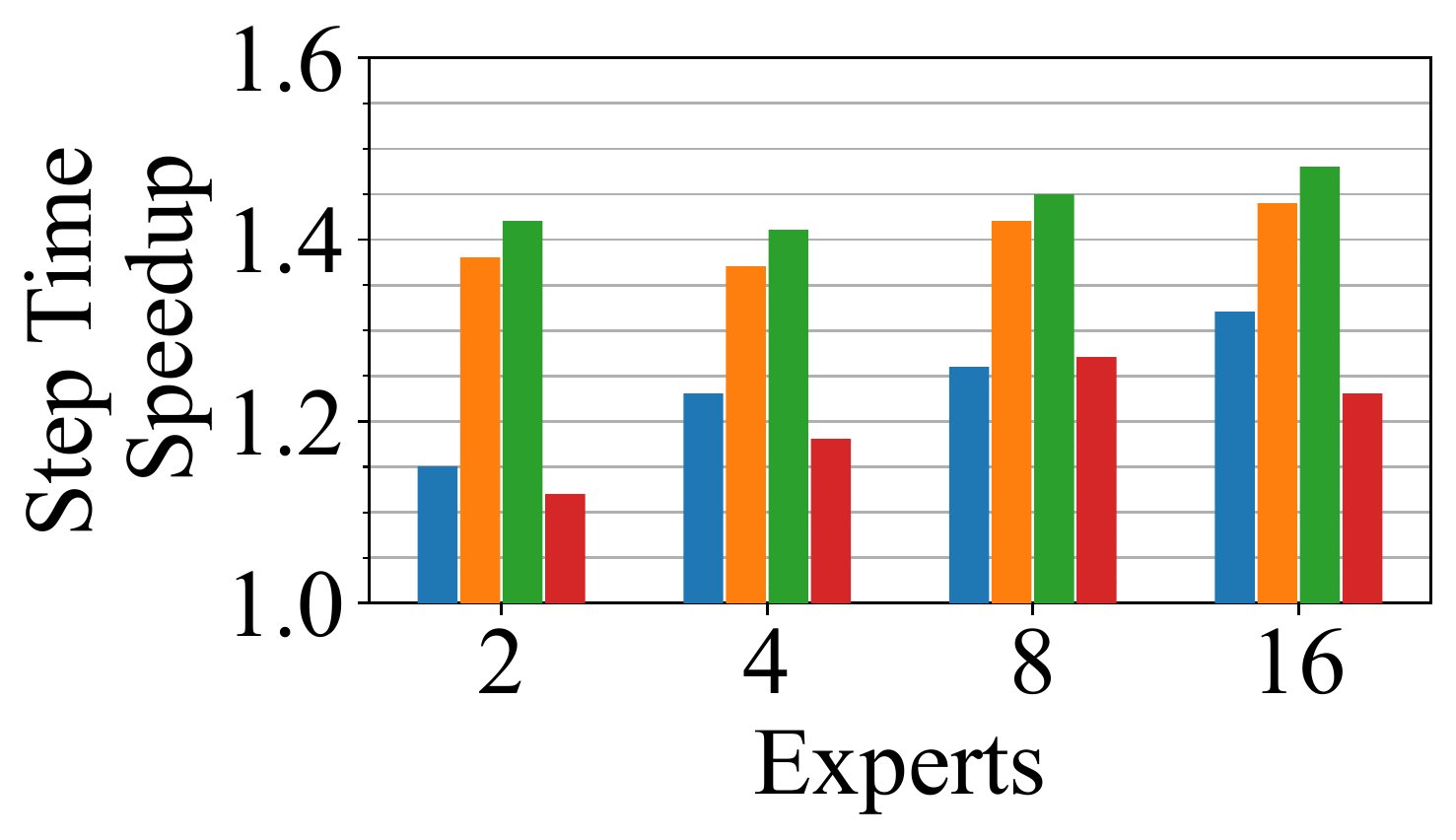}
        \caption{Transformer-XL.}
        \label{fig:xl_schedule_speedup_compare}
    \end{subfigure}%
    \hfill
    \begin{subfigure}{0.33\linewidth}
        \centering
        \includegraphics[width=\linewidth]{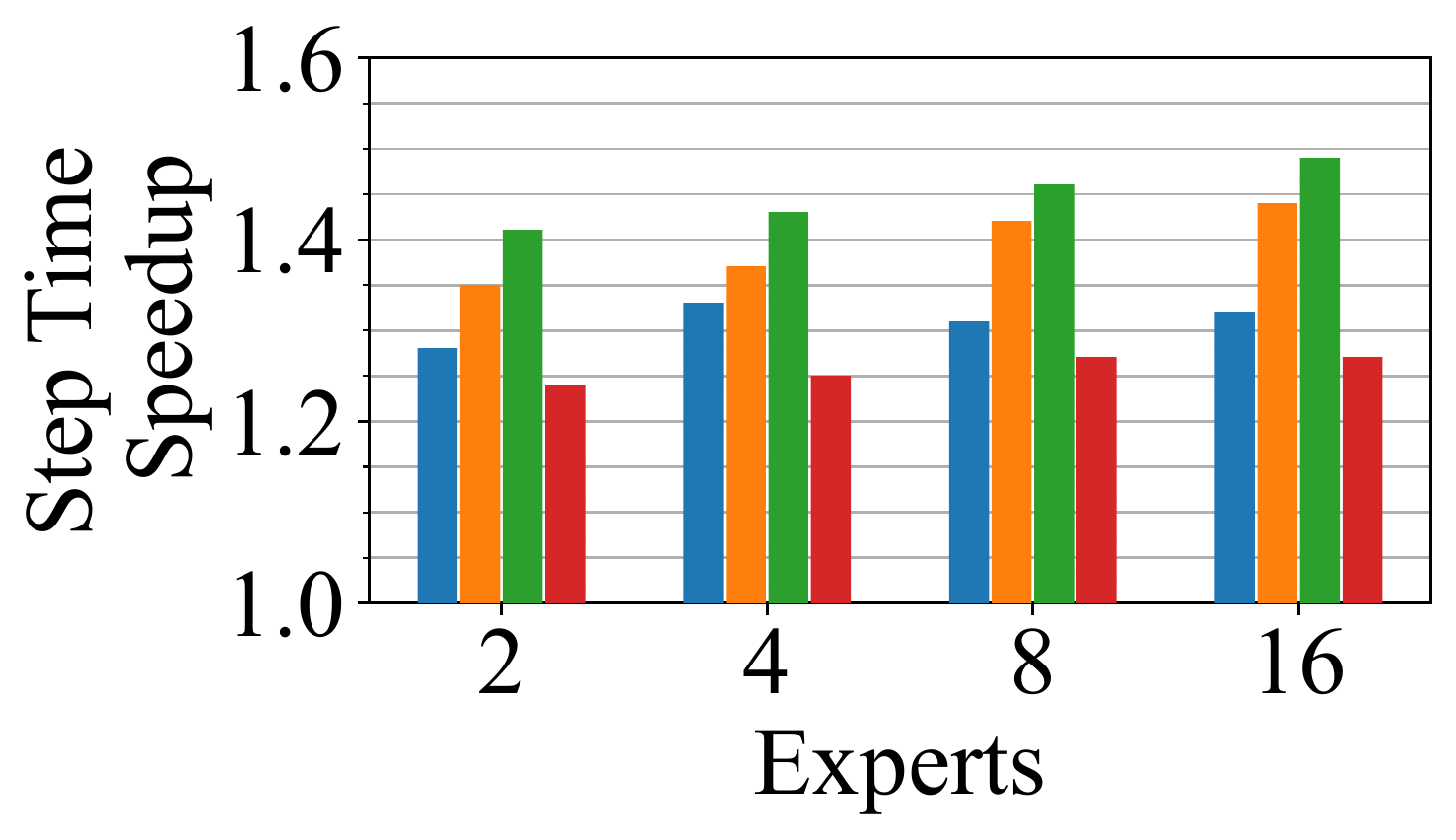}
        \caption{GPT-2. }
        \label{fig:gpt_schedule_speedup_compare}
    \end{subfigure}%
    \hfill
    \begin{subfigure}{0.33\linewidth}
        \centering
        \includegraphics[width=\linewidth]{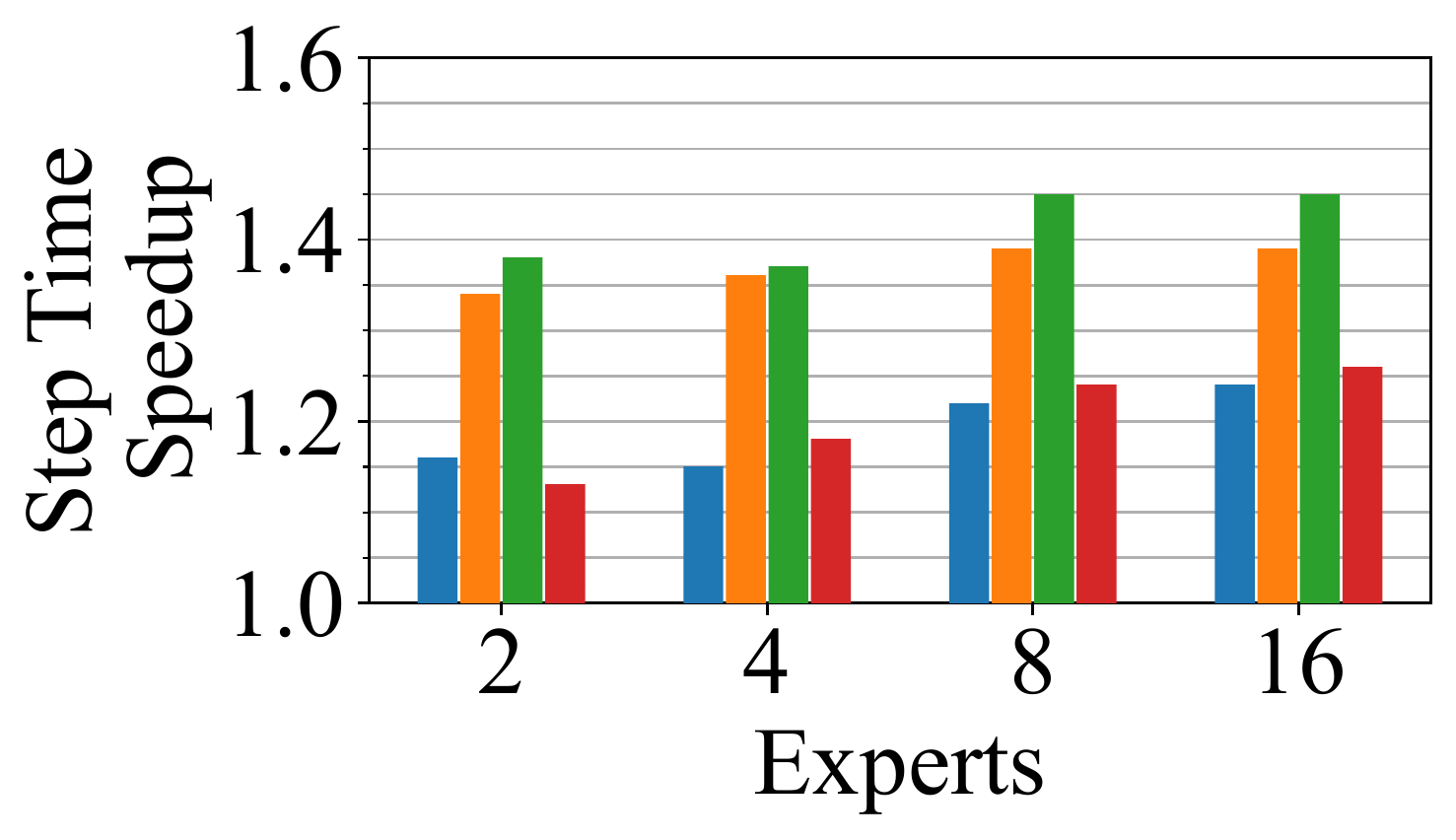}
        \caption{BERT2GPT2.}
        \label{fig:bertgpt_schedule_speedup_compare}
    \end{subfigure}
    \vspace{-2mm}
    \caption{Training step time speedup over Baseline with different design choices of the communication scheduler.}
    \label{fig:schedule_speedup_compare}
\end{minipage}% 
\hfill
\begin{minipage}{0.25\textwidth}
    \centering
    \includegraphics[width=\linewidth]{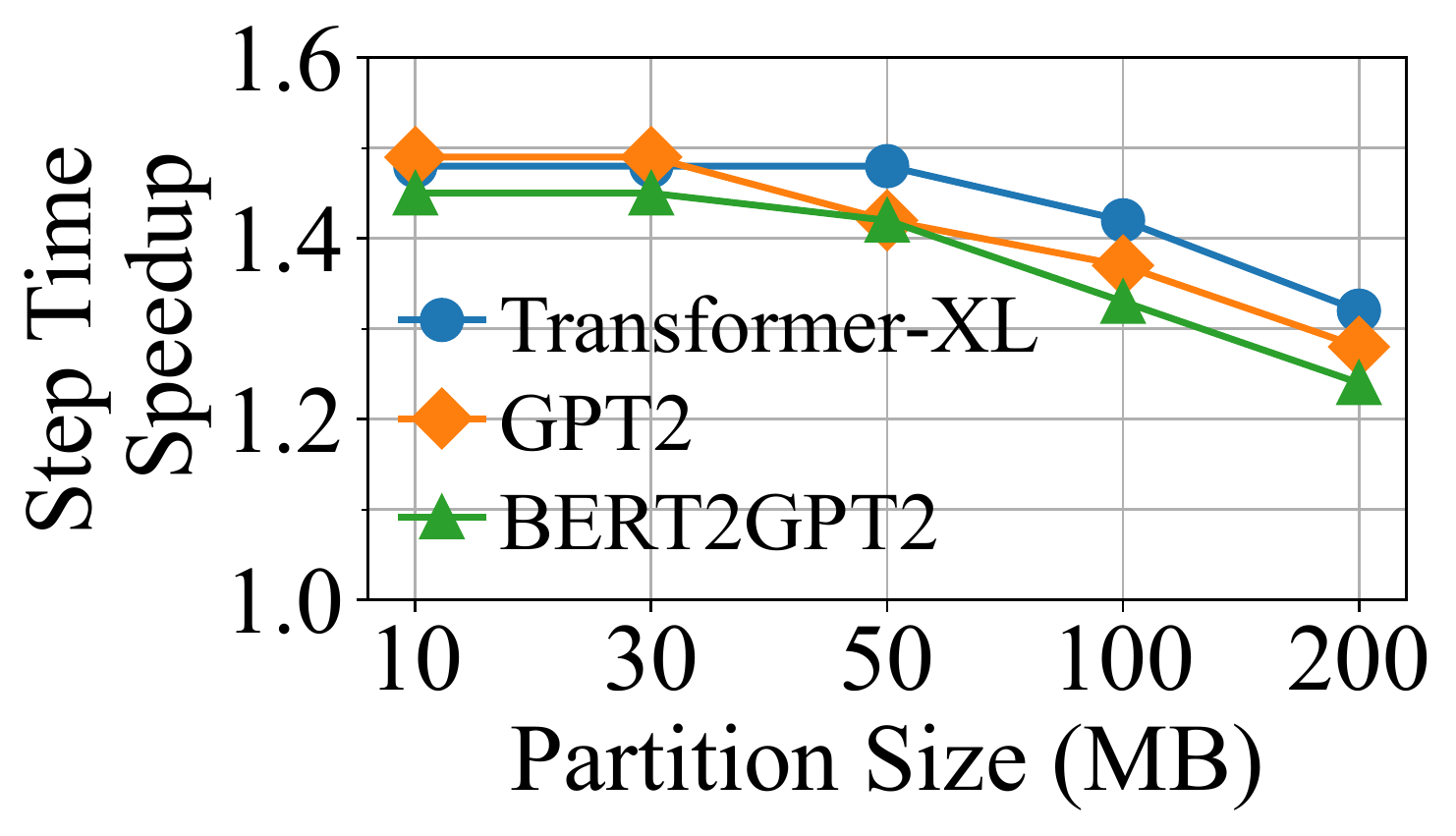}
    \caption{Partition size increases from 10MB to 200MB in 16-expert models.}
    \label{fig:partitionsize_schedule}
\end{minipage}
    \vspace{-4mm}
\end{figure*}
\noindent\textbf{Metrics.}
We consider four metrics to evaluate \sys. 
\begin{itemize}[leftmargin=*]
    \item Training step time: Time to complete one step of training.
    \item Inference time: Time to complete one batch of inference.
    \item \Op time: The completion time of \op. 
    \item MoE layer time: 
    Time to complete one MoE layer of computation and communication.
\end{itemize}
In collecting these metrics we use PyTorch Profiler to obtain CUDA kernel execution time and GPU activities. Training results are averaged over 50 steps after a 10-step warm-up period.
Inference results are averaged over the test set. 
Since the optimization introduced by \sys does not affect the precision of model parameters, model accuracy is unaffected and we omit its evaluation.

\noindent\textbf{Training configurations.}
\sys's micro-op communication scheduler adopts a tensor partition size of 30MB, which can minimize the period blocked by \op in most cases. Expert packing is launched at the 10-th step of each training task and is adjusted every four steps. 

\noindent\textbf{Inference configurations.} \sys's resource scheduler runs on device 0. The path length $l$ in popularity estimation is 3; the maximum number of experts packed on a device is 4. 

\noindent\textbf{Baselines.}
We use the vanilla DeepSpeed~\cite{deepspeed} as the Baseline.
We also provide a comparison to the open-source version of Tutel~\cite{tutel}, which performs similarly with DeepSpeed. \revise{We enable hierarchical \op for both \sys and DeepSpeed and disable Random Token Dropping~\cite{yao2022random} introduced by DeepSpeed.}

\subsection{Training} 
\label{subsec:training_perf}
We start with \sys's training performance. 
Note that \sys is evaluated when the expert packing decision is stabilized; all settings here use 2 experts per device as the best strategy except Transformer-XL with 16 experts, which uses 4 experts per device.
The number of GPUs is equal to the number of experts per layer in both Baseline and \sys. 

\subsubsection{Overall Performance}
\noindent\textbf{Training step time.}
Figure~\ref{fig:overall_step_speedup} shows \sys's speedup in step time over Baseline and Tutel. 
All other aspects of the models stay the same (e.g. sequence length, hidden states dimension, etc.).
Compared to Baseline (DeepSpeed), step time is reduced by an average of 1.37x and 1.47x for the 4- and 16-expert cases, respectively, and by an average of 1.71x and 1.73x for 2- and 8-expert models, respectively.  
The 2- and 8-expert cases see more significant gains as \sys's packs two experts per device as mentioned before. 
The 2-expert case thus boils down to pure data parallelism without any \op; the 8-expert models avoid inter-node \op as our servers have 4 GPUs each. 
\sys's speedup over Tutel is slightly smaller than that of DeepSpeed. Thus in the following we only use DeepSpeed as the Baseline. 

\begin{comment}
\noindent\textbf{MoE layer time and GPU.}
We specifically seek to understand \sys's gain in MoE layers in both the forward and backward pass.
As Figures~\ref{fig:overall_moefwd_speedup} and \ref{fig:overall_moebwd_speedup} show, similar to step time, the gain in the 2- and 8-expert cases is the largest. 
The forward and backward pass of MoE layers in the 2-expert case are accelerated by 1.84x and 2.41x, and in the 8-expert case by 1.89x and 2.32x, respectively.
Meanwhile, since backward pass in Baseline suffers from the interference of \ar while the forward pass does not, the improvement in the backward pass is more significant.
\end{comment}
\noindent\textbf{MoE layer time.}
We specifically seek to understand \sys's gain in MoE layers in both the forward and backward pass.
As Figures~\ref{fig:overall_moefwd_speedup} and~\ref{fig:overall_moebwd_speedup} show, similar to step time, the gain in the 2- and 8-expert cases is the largest. 
The forward and backward pass of MoE layers in the 2-expert case are accelerated by 1.84x and 2.41x, and in the 8-expert case by 1.89x and 2.32x, respectively. Since backward pass in Baseline suffers from the interference of \ar while the forward pass does not, the improvement in the backward pass is more significant. Average GPU utilization in the MoE layer for 16-expert cases is improved by at least 16\% as the period blocked by \op is minimized with \sys.

\revise{
\noindent\textbf{GPU utilization and memory usage.} We measure the average GPU utilization GPU memory usage. We observe an average of 17.6\% improvement in GPU utilization due to the efficient scheduling of \sys. Expert packing would lead to usage increase in GPU memory. The peak memory of BERT2GPT2 is increased by 19.5\% while Transformer-XL and GPT-2 use up all the memory and apply DRAM-offloading to store the packed expert parameters. 
}

\noindent\textbf{\Op time.}
We then zoom in on \op time in backward pass, where \sys prioritizes \op and avoids concurrent execution with \ar. 
Expert packing also reduces the \op transfer size. 
Figure~\ref{fig:overall_a2a_speedup} shows an average speedup 
of 2.21x, 2.39x, and 2.31x in 4-, 8-, and 16-expert cases in \op time, respectively. 

We also examine the pipelining efficiency between \op and expert computation in \sys. 
We define the pipelining efficiency to be the fraction of non-idle time in the computation CUDA stream during the \op duration. We calculate the pipelining efficiency of \sys before and after adopting expert packing in Table~\ref{table:pp_eff_eval}.
The average improvement is 2.43x in 16-expert case, which also demonstrates the benefits of expert packing. 
The expert FFN micro-op time is thus closer to the \op time. 
We find that two experts per device can achieve the best pipelining efficiency in most cases, justifying our settings mentioned before.

\begin{table}[t] 
    \centering
    \resizebox{0.85\columnwidth}{!}{
        \begin{tabular}{@{}clccc@{}}
            \toprule
            \multirow{2}{*}{Expert} & \multirow{2}{*}{Model} & \multicolumn{3}{c}{Pipelining Efficiency} \\ \cmidrule(l){3-5} 
            &    & \multicolumn{1}{c}{w/o Packing} & \multicolumn{2}{c}{w/ Packing (Experts per Device)} \\ \midrule
            \multirow{3}{*}{16} & Transformer-XL  & 33\%   & 86\%    & 4   \\
                                & GPT-2           & 36\%   & 85\%    & 2   \\
                                & BERTGPT2        & 34\%   & 79\%    & 2   \\ \bottomrule
            \end{tabular}}
            \vspace{-2mm}
    \captionof{table}{Pipelining efficiency comparison with and without expert packing.}
    \vspace{-3mm}
    \label{table:pp_eff_eval}
\end{table}

\begin{table}[t] 
    \centering
    \resizebox{\columnwidth}{!}{
        \begin{tabular}{@{}clccccc@{}}
            \toprule
            \multirow{2}{*}{Expert} & \multirow{2}{*}{Model} & \multicolumn{2}{c}{Average GPU Utilization(\%)} & \multicolumn{3}{c}{GPU Memory Peak Usage(\%)} \\ \cmidrule(l){3-7} 
             &  & Baseline & \sys & Baseline & \sys & DRAM-offloading \\ \midrule
             \multirow{3}{*}{16} & Transformer-XL & 66.2 & 83.4 & 72.1 &  100 & \cmark \\
             & GPT2 & 62.3 & 78.2 & 83.8 & 100 & \cmark \\
             & BERT2GPT2 & 63.5 & 82.5 & 74.3 & 94.2 & \xmark \\ \bottomrule
            \end{tabular}}
            \vspace{-2mm}
    \captionof{table}{\revise{GPU utilization and peak memory usage of 16-expert MoE models. GPU Memory Peak Usage is the ratio between the maximum usage and the total device memory. DRAM-offloading indicates if it is applied. }}
    \vspace{-6mm}
    \label{table:gpu_util_mem}
\end{table}

\subsubsection{Communication Scheduler}
\label{subsec:scheduler_perf}
We now present an in-depth analysis of \sys's priority-based micro-op scheduler, aiming to understand the benefit of each design choice.
For fairness all experiments here are obtained without expert packing in \sys, i.e. one expert per device.
\begin{figure*}
    \centering
    \includegraphics[width=0.65\linewidth]{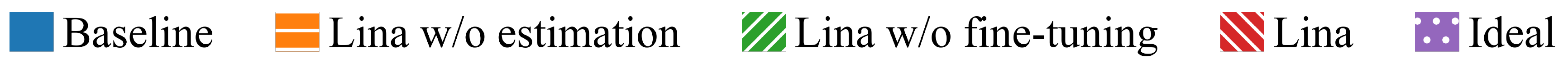}
    \vspace{-3mm}
\end{figure*}
\begin{figure*}
    \begin{subfigure}{0.24\textwidth}
        \centering
        \includegraphics[width=\linewidth]{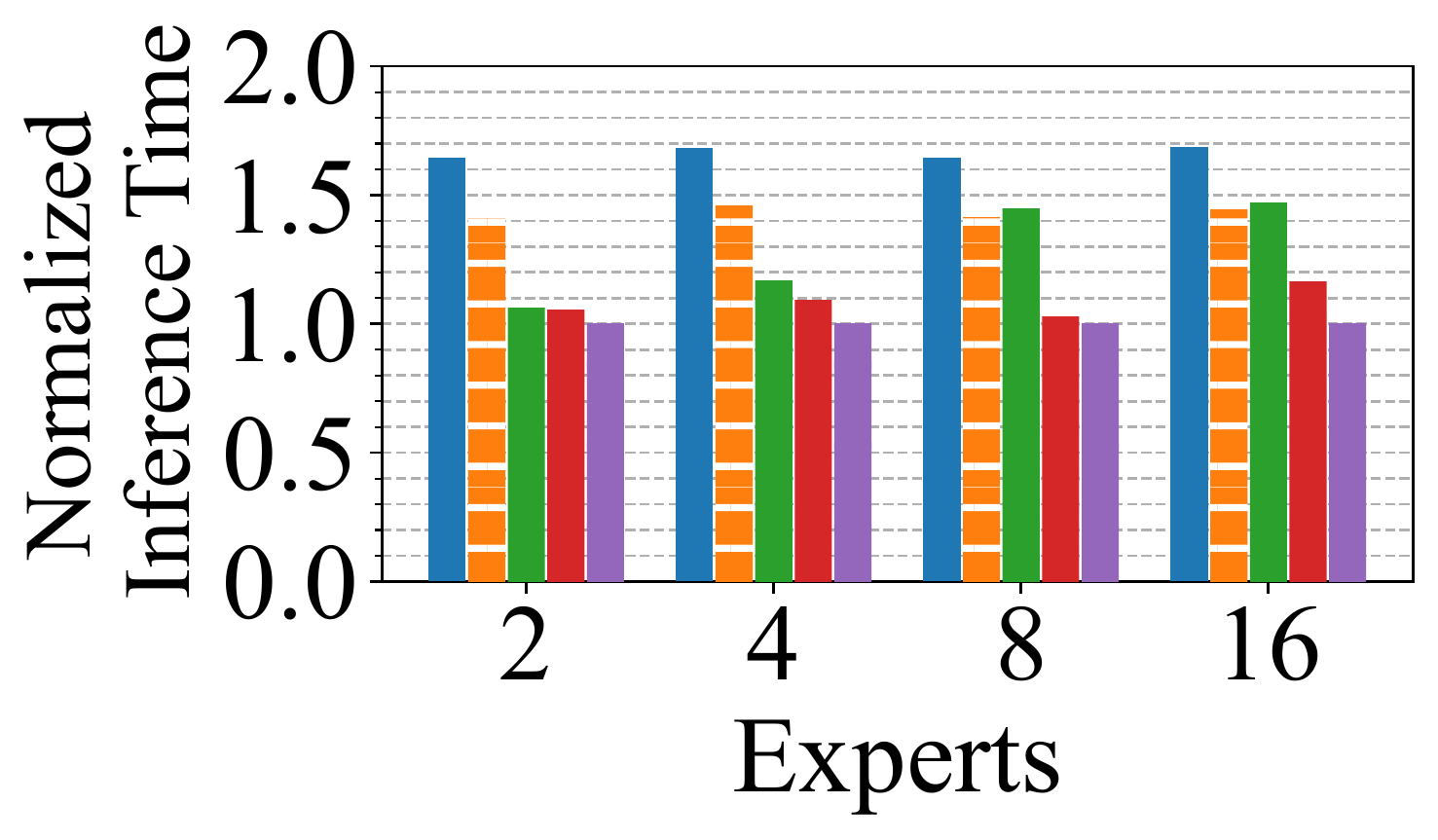}
        \vspace{-4mm}
        \caption{Median inference time with Transformer-XL.}
        \label{fig:transformer_latency_speedup}
    \end{subfigure}%
    \hfill
    \begin{subfigure}{0.24\textwidth}
        \centering
        \includegraphics[width=\linewidth]{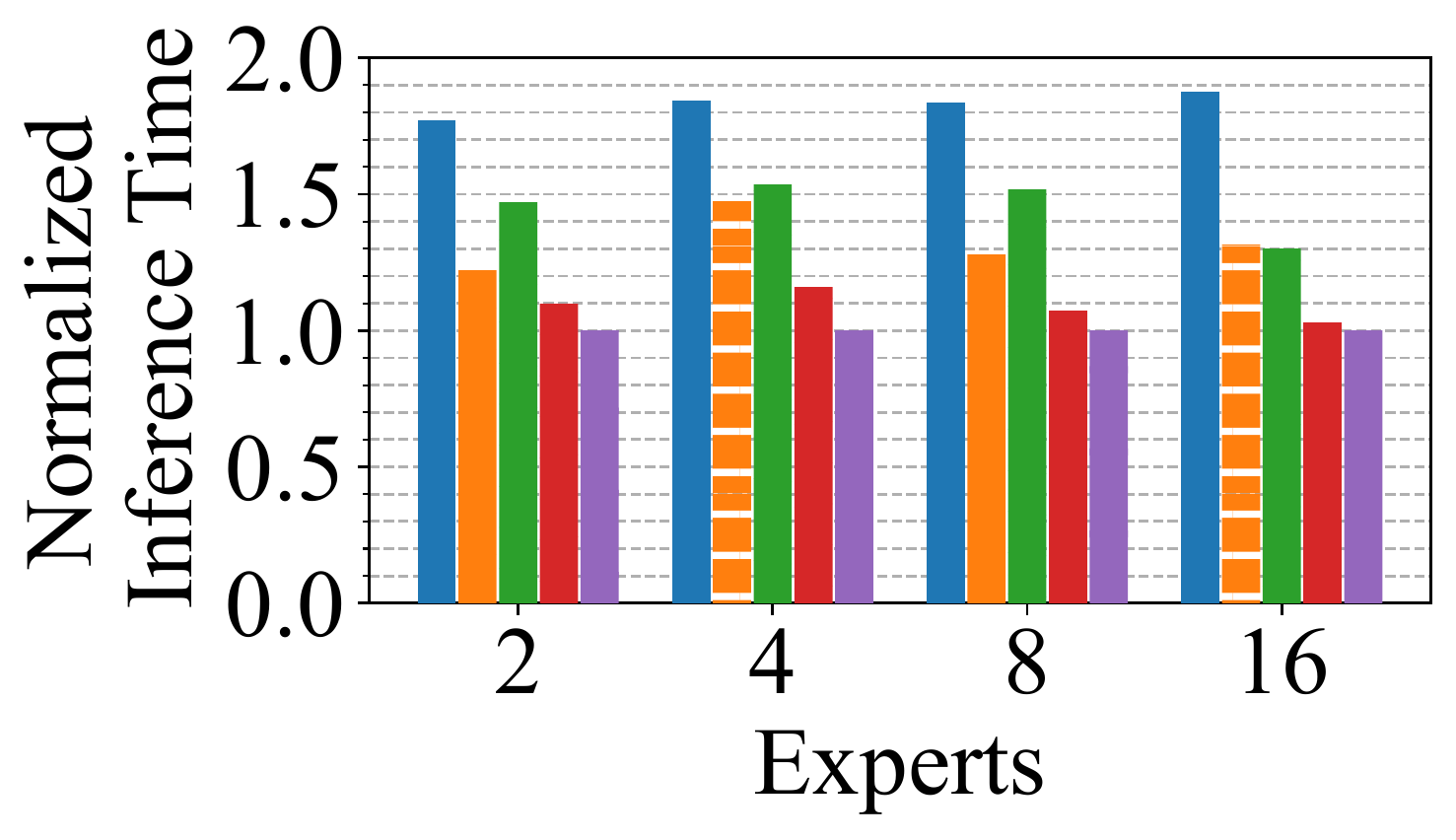}
        \vspace{-4mm}
        \caption{95\%ile inference time with Transformer-XL.}
        \label{fig:transformer_tail_latency_speedup}
    \end{subfigure}%
    \hfill
    \begin{subfigure}{0.24\textwidth}
        \centering
        \includegraphics[width=\linewidth]{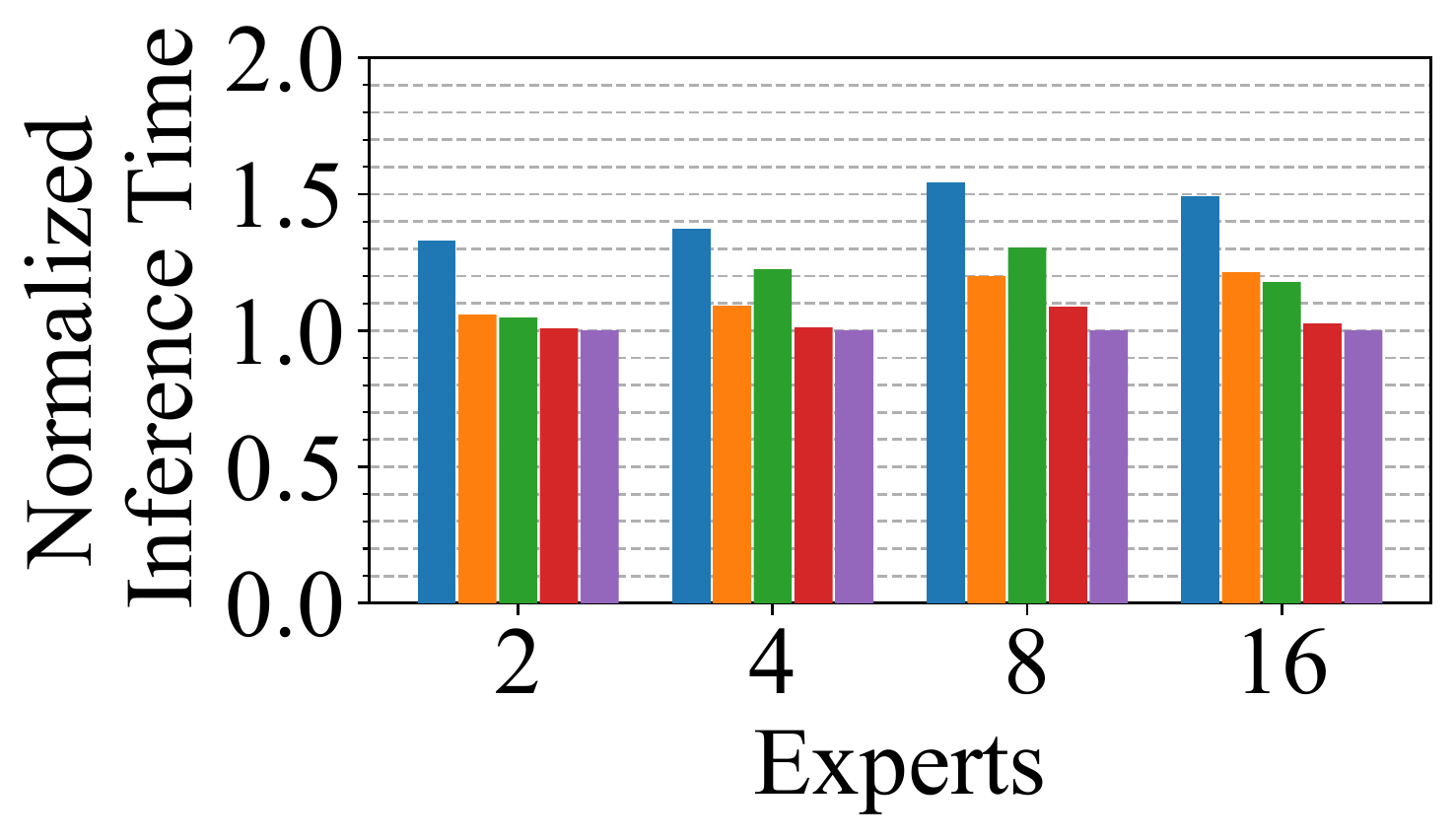}
        \vspace{-4mm}
        \caption{Median inference time with BERT-Large.}
        \label{fig:bert_latency_speedup}
    \end{subfigure}%
    \hfill
    \begin{subfigure}{0.24\textwidth}
        \centering
        \includegraphics[width=\linewidth]{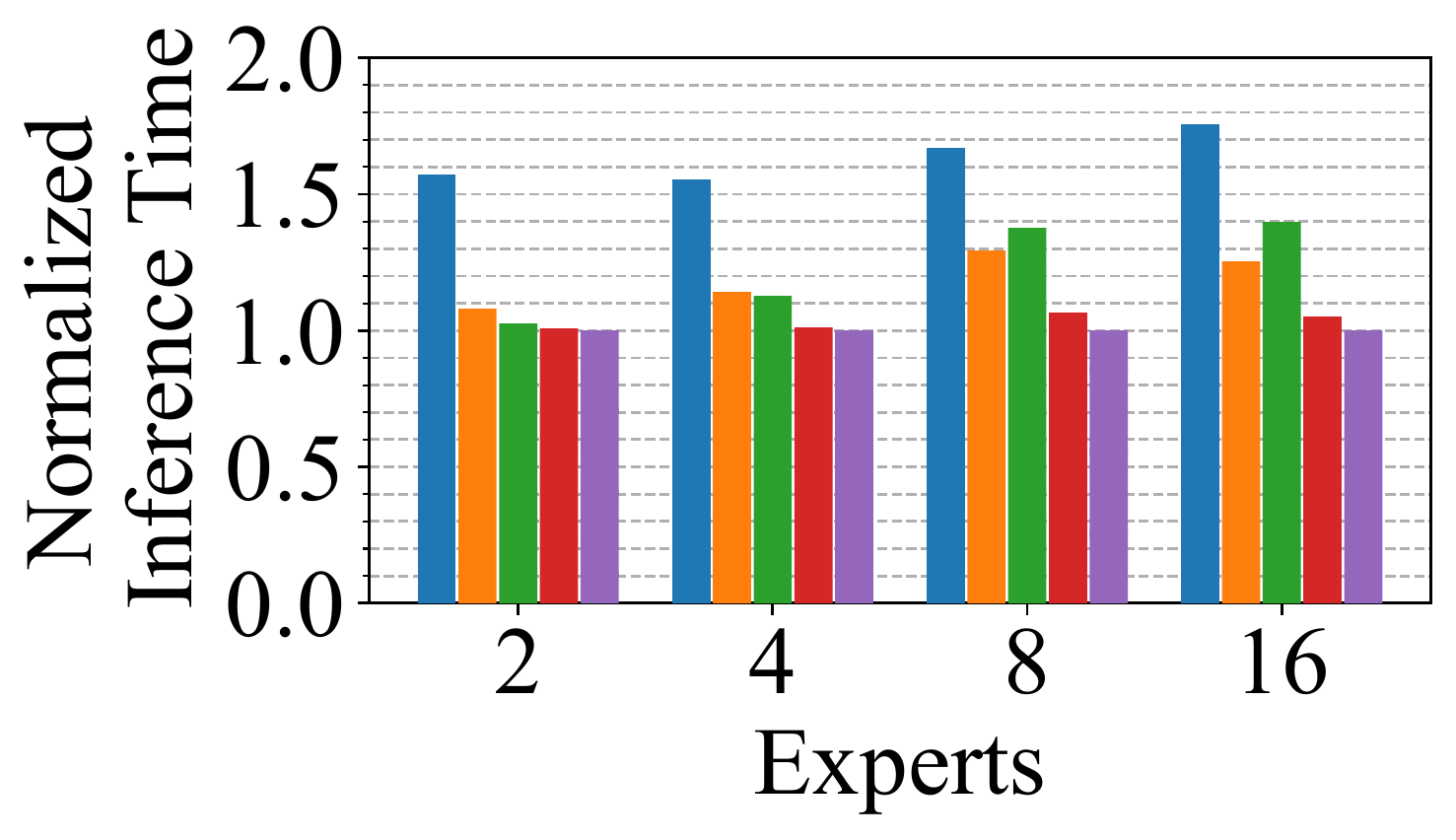}
        \vspace{-4mm}
        \caption{95\%ile inference time with BERT-Large.}
        \label{fig:bert_tail_latency_speedup}
    \end{subfigure}%
    \vspace{-2mm}
    \caption{Median and tail inference time. We normalize the inference time with the ideal result. The median and tail inference time is the same in Ideal. }
    \label{fig:avg_tail_latency}
    \vspace{-5mm}
\end{figure*}
\noindent\textbf{Tensor partitioning and pipelining.}
To justify our design, we incrementally add the key design choices to Baseline and see their corresponding gain: first priority scheduling, then tensor partitioning, and lastly pipelining. 
Besides, we consider a fixed scheduling strategy where \ar is always scheduled between pairs of \op operations (i.e. two MoE layers) with tensor fusion enabled in PyTorch's \texttt{DistributedDataParallel} by default (same as Baseline).

Figure~\ref{fig:schedule_speedup_compare} shows the step time comparison. 
We make several interesting observations here. 
First, using priority brings about 10\%--30\% gain over Baseline in most cases, with an average of 24\%. 
Priority scheduling in general presents more benefit when more devices and nodes are used in training. 
The main reason is that \op's slowdown due to sharing bandwidth with \ar is more severe as training scales out.
Second, tensor partitioning significantly improves the benefit of prioritizing \op: step time is reduced over Baseline by 
1.36x, 1.36x, 1.41x and 1.42x in 2-, 4-, 8-, and 16-expert cases, respectively on average. 
On the other hand, pipelining's gain is limited as expected, since expert computation takes much less time than \op without expert packing (recall \cref{subsec:schedule_design}).
Overall, all three design choices can effectively reduce \op's completion time.

We also observe that the relative benefit of priority scheduling and tensor partitioning is model-specific: GPT-2 enjoys much more gain from priority compared to tensor partitioning while the other two models do not exhibit such clear pattern. 
This is likely due to the degree of overlapping of \op and \ar: most \ar can fit in between \op operations  in GPT-2, and as a result using priority scheduling alone is very beneficial.

Finally, the fixed scheduling strategy leads to the smallest gains in almost all cases. 
This is because (1) \op still has to fair-share bandwidth with \ar, and (2) tensors are not partitioned which aggravates the impact of \ar.
This demonstrates again the effectiveness of our design in prioritizing \op with smaller tensors instead of using fixed heuristics that cannot opportunistically maximize efficiency.

\noindent\textbf{Partition size.} 
We also evaluate the impact of partition 
size on the communication scheduler. 
Figure~\ref{fig:partitionsize_schedule} shows the step time of 16-expert 
models when we gradually increase the partition size from 10MB to 100MB. 
We find that a partition size beyond 50MB slows down Transformer-XL and BERT2GPT2 compared with 30 MB. 
As long as the period blocked by \op is minimized, step time would be minimum.  
Therefore, for each model and setting, there are multiple optimal partition sizes.
Ideally, the scheduler can more precisely control the operations with a smaller partition size. In practice, small partitions (below 10MB) may cause heavy transmission overhead in each micro-op and degrade the overall performance~\cite{bytescheduler}.

\noindent\textbf{Overhead analysis.}
We provide a brief analysis of the overhead incurred by \sys's communication scheduler. First, the preprocessing 
and postprocessing, including tensor partitioning and concatenation, take an average 1.02\% of the step time. Second, 
we measure the transmission overhead of micro-ops. We sum up running times of all the communication micro-ops and compare against those without partitioning in Baseline. 
The average completion time is lengthened by 1.7\%.%, which is negligible.  

\begin{figure*}
    \centering
    \includegraphics[width=0.25\linewidth]{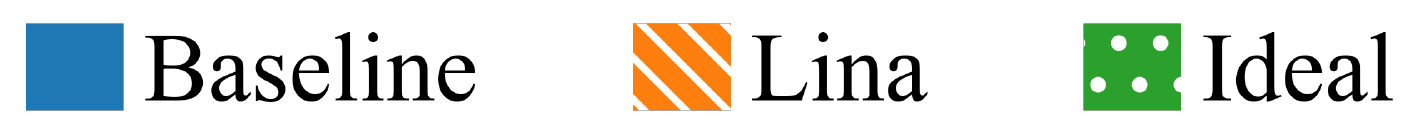}
    \vspace{-3mm}
\end{figure*}
\begin{figure*}
    \begin{minipage}{0.48\textwidth}
    \begin{subfigure}{0.49\textwidth}
        \centering
        \includegraphics[width=\linewidth]{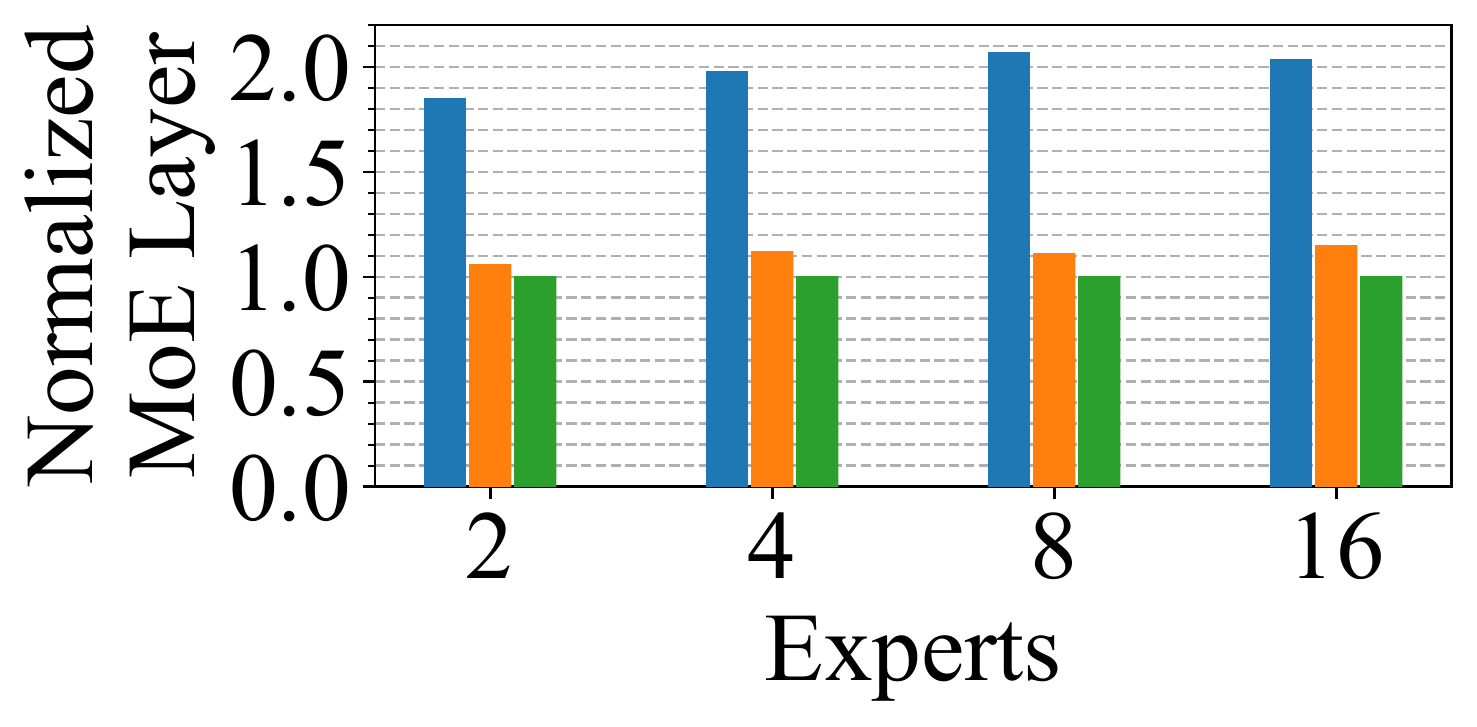}
        \vspace{-5mm}
        \caption{Transformer-XL.}
        \label{fig:transformer_tail_moe_latency_speedup}
    \end{subfigure}% 
    \hfill
    \begin{subfigure}{0.49\textwidth}
        \centering
        \includegraphics[width=\linewidth]{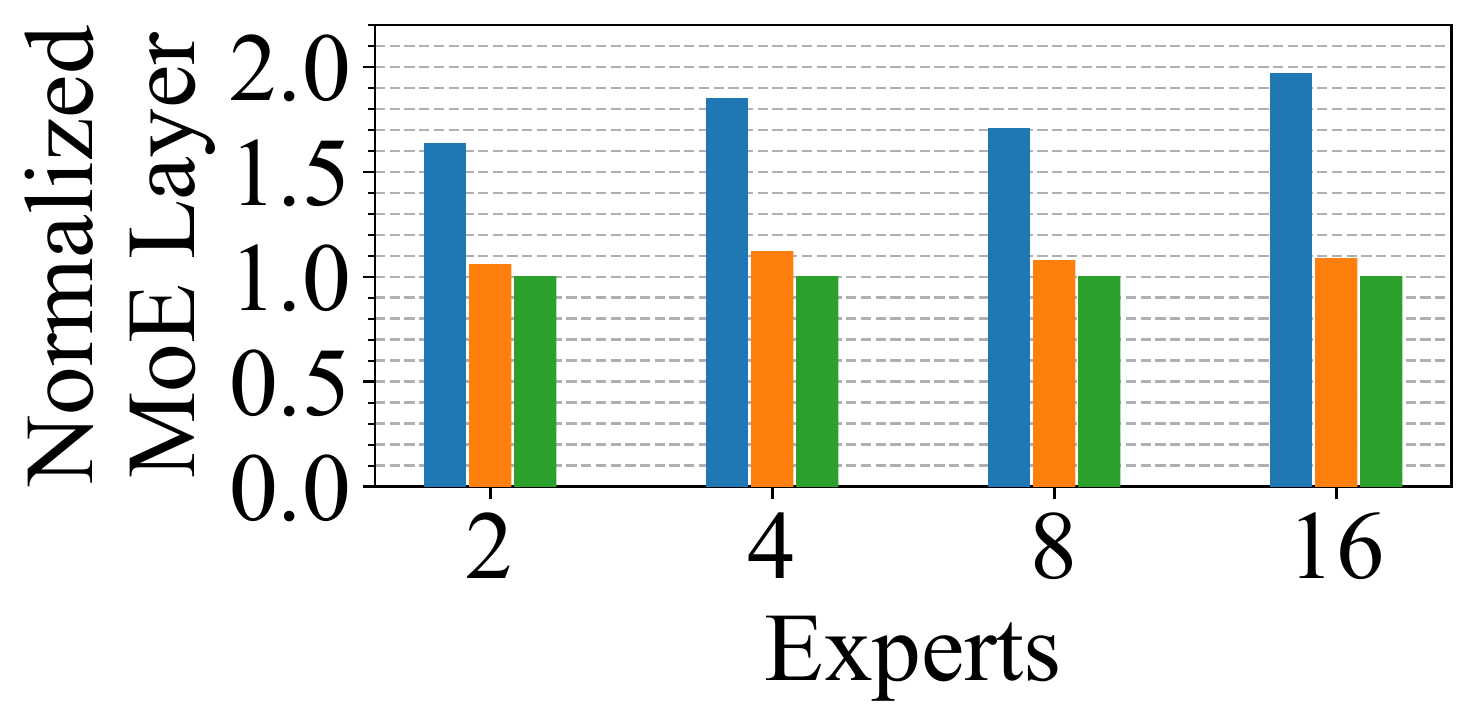}
        \vspace{-5mm}
        \caption{BERT-Large.}
        \label{fig:bert_tail_moe_latency_speedup}
    \end{subfigure}
    \vspace{-3mm}
    \caption{95\%ile completion time of MoE layer.}
    \label{fig:tail_moe_latency_speedup}
    \end{minipage}% 
    \hfill
    \begin{minipage}{0.28\textwidth}
        \centering
        \includegraphics[width=\linewidth]{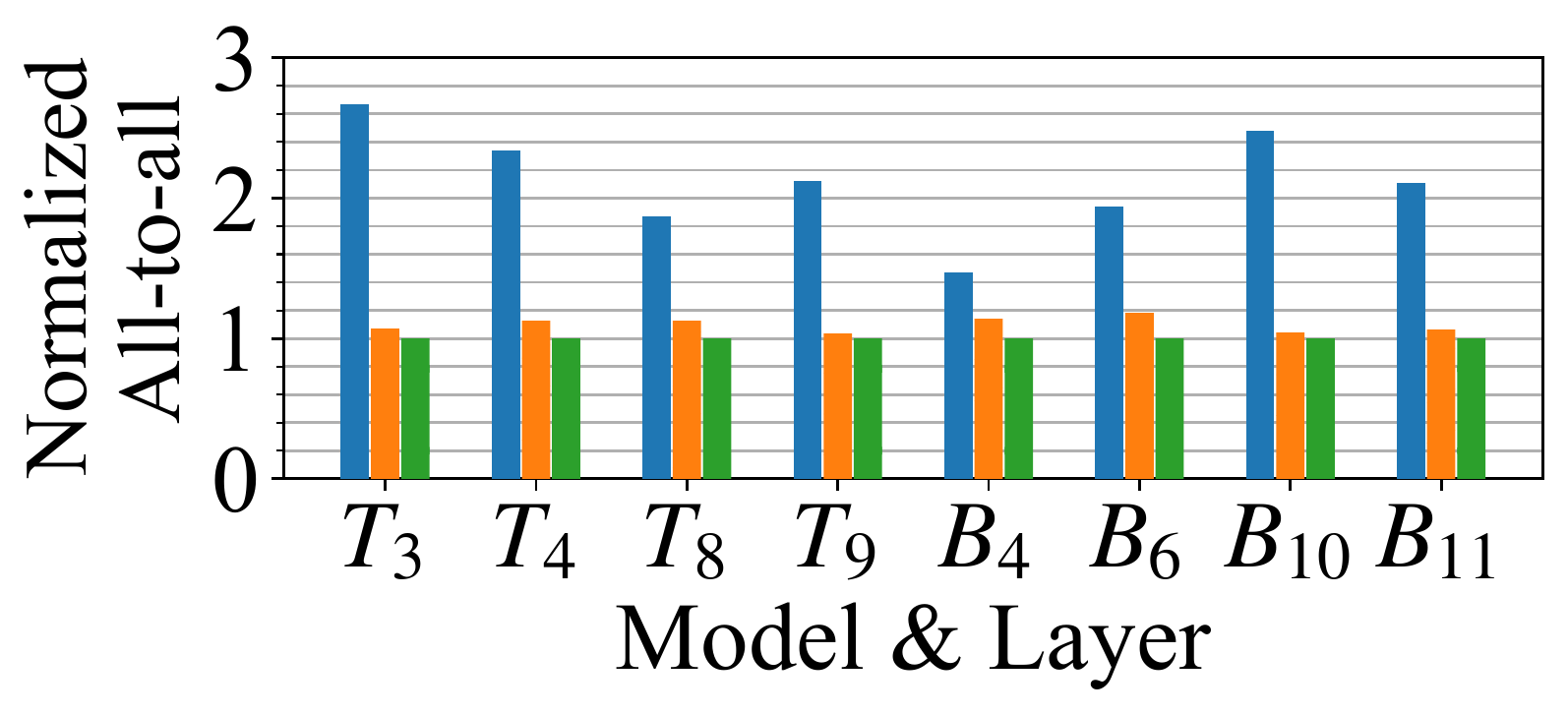}
        \vspace{-8mm}
        \caption{\Op time in 16-expert MoE. $T$ is Transformer-XL and $B$ is BERT-Large.}
        \label{fig:inference_a2a_latency}
    \end{minipage}% 
    \hfill
    \begin{minipage}{0.23\textwidth}
        \centering
        \includegraphics[width=\linewidth]{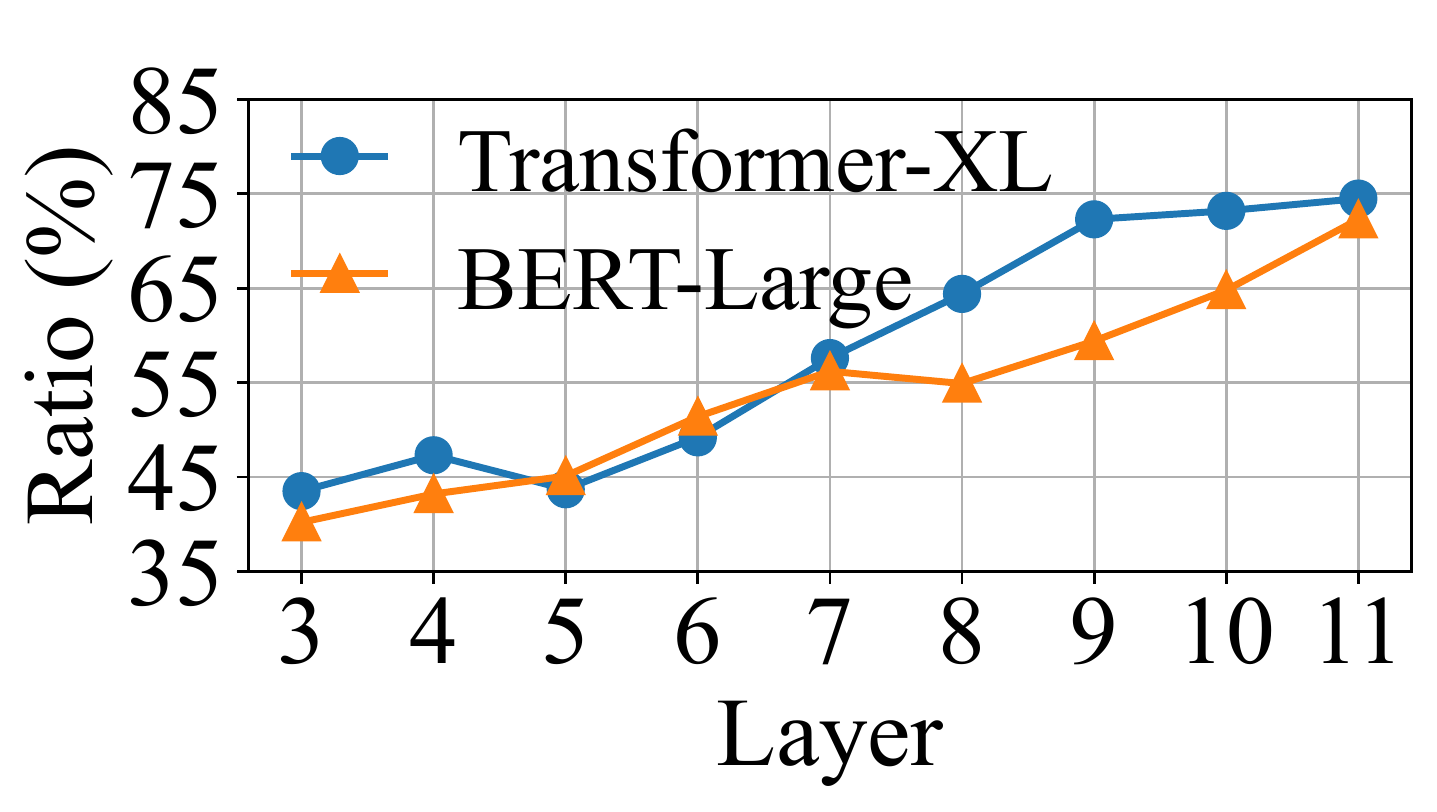}
        \vspace{-8mm}
        \caption{Estimation accuracy of 16-expert MoE.}
        \label{fig:popularity_estimation_accuracy}
    \end{minipage}
    \vspace{-3mm}
\end{figure*}

\subsection{Inference}
\label{subsec:inference_perf}
We then evaluate \sys's inference performance. 
Each experiment is repeated five times: two of which measure the end-to-end inference time, and the rest profile the different components with Profiler and collect statistics for overhead and estimation accuracy. This way the inference time is not affected by the profiling overhead.

\subsubsection{Resource Scheduler}
\label{subsubsec:inference_sched_perf}
\noindent\textbf{Inference time.} 
Figure~\ref{fig:avg_tail_latency} shows the median and 95\% inference time of Baseline and \sys. We also present the ideal inference time with a perfectly balanced load across devices in all MoE layers. This is obviously challenging to achieve with real-world requests. 
Thus we modify the gating network to constantly output a balanced expert selection to obtain this benchmark. 
We normalize all results to the Ideal value. 

\sys's resource scheduler effectively balances the load among devices and achieve inference time close to Ideal. 
Compared to Baseline, median inference time is reduced by 1.54x and 1.45x for the 4- and 16-expert Transformer-XL, and by 1.36x and 1.46x for the 4- and 16-expert BERT-Large, respectively. 
The 95\%ile inference time is reduced by 1.82x for 16-expert Transformer-XL and 1.68x for 16-expert BERT-Large. 
The reduction on tail inference time increases with more experts in a layer, because a wider MoE layer is more likely to present more skewed expert popularity, giving more room for \sys to optimize. 
Lastly, \sys's gap to Ideal can be explained for two reasons other than the overheads. 
First, \sys cannot perfectly balance load: the least popular experts are randomly placed for example. 
Second, 
\sys starts to schedule from the fourth layer.  

\noindent\textbf{MoE layer and \op time.} 
With \sys, MoE layer time includes gate computation, phase two of scheduling, two \op, and expert computation; phase one of the scheduling is largely overlapped with computation as explained in \cref{subsec:inference_impl}. 
The 95\%ile MoE layer time is reduced by 1.87x and 1.77x in 8- and 16-expert Transformer-XL over Baseline and by 1.58x and 1.81x in 8- and 16-expert BERT-Large as in Figure~\ref{fig:tail_moe_latency_speedup}. 
We also extract \op time, which is a direct indicator of whether \sys balances load across devices effectively. 
We present the tail \op time reduction of different layers in Figure~\ref{fig:inference_a2a_latency}. 
The average and maximum improvements are 1.96x and 2.50x over Baseline. 
These results confirm that \sys effectively balances the load of each device and \op transfer size of each link.

\noindent\textbf{Two-phase scheduling.} 
We then evaluate the effectiveness of our resource scheduler's design. 
We consider separately \sys without estimation and without fine-tuning in order to understand their individual gains. 
\sys w/o estimation refers to scheduling using the actual routing decision computed by the gating network.

In Figure~\ref{fig:avg_tail_latency}, we present the comparison of inference time for all schemes. 
Without estimation the median inference time is worsened by 24.0\% and 18.6\% for 16-expert Transformer-XL and BERT-Large in \sys. 
The scheduler works after the gating network and blocks \op with the following computation until it completes. 
Thus the scheduling overhead manifests at each MoE layer, overweighing the additional gains brought by accurate popularity information. 
The tail inference time is less affected compared to the median, but still suffers without estimation.  

Without fine-tuning, tail inference time is prolonged by 26.7\% and 33.1\% for 16-expert Transformer-XL and BERT-Large. 
This suggests that fine-tuning also plays an indispensable role when the estimation shows a large difference from the actual routing decision. 
For example, if the top-1 expert in the actual routing decision is estimated as an unpopular one packed with others, the Moe layer time would even be worse than Baseline. 
More discussions are presented in~\Cref{subsubsec:estimation_perf}. The importance of fine-tuning depends heavily on estimation accuracy and number of expert in MoE layer.

\noindent\textbf{Overhead analysis.} 
We dissect the overhead of the resource scheduler, by considering the scheduling times of phase one and phase two separately. 
The scheduling time for both phases averages at $\sim$6.2ms since they share the same logic and coordination workflow. 
Yet, the overhead of phase one is largely overlapped with model computation.
Though overhead of phase two with re-scheduling is more salient, it only kicks in for 23\% of the cases on average, and is smaller than the idle time incurred by skewed expert popularity. 
The overhead of phase two without fine-tuning is merely 1.45ms. 

\begin{table}[t]
    \centering
    \resizebox{0.95\columnwidth}{!}{
        \begin{tabular}{@{}lccccc@{}}
            \toprule
            \multicolumn{1}{l}{\multirow{2}{*}{Model}} & \multirow{2}{*}{\begin{tabular}[c]{@{}c@{}}Path\\ Length\end{tabular}} & \multicolumn{2}{c}{Norm. Inference Time} & \multirow{2}{*}{\begin{tabular}[c]{@{}c@{}}Fine-tuning\\ (\%)\end{tabular}} & \multicolumn{1}{c}{\multirow{2}{*}{\begin{tabular}[c]{@{}c@{}} Estimation\\ Accuracy (\%)\end{tabular}}} \\ 
            \multicolumn{1}{c}{} &  & \multicolumn{1}{c}{Median} & \multicolumn{1}{c}{95\%ile} &  & \multicolumn{1}{c}{} \\ \midrule
             & 1 & 1.41 & 1.32 & 76.5 & 31.6 \\
            Transformer-XL & 3 & 1.16 & 1.04 & 25.7 & 60.4 \\
             & 6 & 1.19 & 1.11 & 22.5 & 71.4 \\ \midrule
             & 1 & 1.34 & 1.35 & 71.3 & 28.3 \\
             BERT-Large & 3 & 1.07 & 1.04 & 32.2 & 63.5 \\
             & 6 & 1.09 & 1.11 & 27.1 & 66.0 \\ \bottomrule
            \end{tabular}}
            \vspace{-3mm}
    \captionof{table}{\sys's performance using different path lengths during estimation. Both models have 16 experts per layer. Inference time is normalized to Ideal. }
    \vspace{-3mm}
    \label{table:path_length}
\end{table}
\subsubsection{Popularity Estimation}
\label{subsubsec:estimation_perf}
We now analyze \sys's popularity estimation method. 

\noindent\textbf{Estimation accuracy.} 
We first examine the estimation accuracy across MoE layers. 
We resort to the same definition used in \sys's phase two scheduling: if the top-2 (recall $k=1$ in inference) estimated experts are identical to the actual routing decision, we consider the estimation accurate. 
Figure~\ref{fig:popularity_estimation_accuracy} shows the accuracy for every MoE layer in two inference tasks. 
Overall, estimation accuracy is 58.41\% and 54.16\% for Transformer-XL and BERT-Large, respectively. 
The estimation accuracy is higher in the latter layers of the model, which is consistent with our observation in Figure~\ref{fig:popularity_pattern}. 
We also compare the complete popularity rankings given by the estimation to better understand its accuracy. 
Using 1000 random batches of the Transformer-XL model, we observe that errors usually happen at experts with a similar popularity.  An average of 3.67 experts out of the estimation are incorrectly ordered. Therefore, the fine-tuning only requires little adjustment to the experts packed together. The effectiveness of \sys's estimation can be justified.

\noindent\textbf{Sample path length.} 
We also investigate the impact of sample path length $l$. 
The longer the sample path of expert selection is for making an estimation, the more accurate the result is. 
Table~\ref{table:path_length} shows \sys's performance degrades with $l=1$, in terms of inference time, estimation accuracy, and the occurrence of phase two fine-tuning, compared with the default length of 3. 
Longer paths can elevate the estimation accuracy and further reduce the number of times of \sys's fine-tuning. However, due to the problem of a slower start, the reduction of inference time is not as noteworthy as the estimation accuracy. For a path length of 6, \sys shows a similar median and tail result as the performance with a path length of 3.

\begin{table}[t]
    \centering
    \resizebox{\columnwidth}{!}{
        \begin{tabular}{@{}lllcc@{}}
            \toprule
            Task & Dataset & Model & Norm. 95\%ile Inference Time & Estimation Accuracy \\ \midrule
            \multirow{2}{*}{\begin{tabular}[c]{@{}l@{}}Sentiment\\ Analysis\end{tabular}} & IMDB Reviews~\cite{imdb} & \multirow{2}{*}{BERT} & 1.08 & 64.4\% \\
             & Twitter~\cite{paws2019naacl} &  & 1.11 & 62.3\% \\ \midrule
            \multirow{2}{*}{\begin{tabular}[c]{@{}l@{}}Translation\\ (English)\end{tabular}} & WMT French~\cite{wmt} & \multirow{2}{*}{T5~\cite{2020t5}} & 1.04 & 68.8\% \\
             & WMT Russian~\cite{wmt} &  & 1.08 & 62.5\% \\ \bottomrule
            \end{tabular}}
            \vspace{-3mm}
    \captionof{table}{\revise{\sys's performance on different tasks and datasets. Inference time is normalized to Ideal. The path length is set to 3.}}
    \vspace{-8mm}
    \label{table:generality_estimation}
\end{table}

\revise{
\noindent\textbf{Generalizability.} We proceed to evaluate how well \sys's popularity estimation approach can be generalized to different tasks. Table~\ref{table:generality_estimation} shows the estimation accuracy of four tasks with different datasets. The 95\%ile inference time can achieve at most 1.04x of the Ideal inference time and the estimation accuracy is at least 62.3\%. \sys's estimation method relies on the patterns obtained from training stage. Therefore, it is tailored to each specific task and proves to be an effective approach to capturing the expert popularity prior.
}

%% file: discussion.tex
\revise{
\section{Discussion}
\label{sec:discussion}
\noindent\textbf{Parallelism in training.} With the increasing scale of language models, the adoption of pipeline and tensor parallelisms has become essential~\cite{shoeybi2019megatron}. Pipeline parallelism involves the use of blocking send and receive operations to transmit intermediate activations, while tensor parallelism utilizes blocking all-reduce operations to combine tensor partitions. Extensive research has been conducted on coordinating communication operations for dense models~\cite{zheng2022alpa, jia2019beyond, tutel}. \sys focuses on sparsely-activated MoE with data and expert parallelism, which are orthogonal to existing work.

\noindent\textbf{Estimation of expert popularity.} The current estimation approach used by \sys relies on data collection during the training stage. While fine-tuning can assist in improving efficient expert placement decisions, an estimation method with improved accuracy and confidence would further reduce inference time. One potential approach is to leverage machine learning techniques to train a compact yet powerful model that can predict the expert selected by each token in every MoE layer ahead of time, when the requests are received. 
}

%% file: related.tex
\section{Related Work}
\label{sec:related}
\noindent\textbf{Existing MoE systems.}
Recent literature has proposed MoE-specific optimization techniques.
DeepSpeed~\cite{rajbhandari2022deepspeedmoe} enables distributed training for MoE models and leverages flexible combinations of parallelism strategies. It also introduces a novel MoE architecture called Pyramid-Residual MoE. PR-MoE applies experts only where they
are most effective. Tutel~\cite{tutel} extends DeepSpeed and proposes an adaptive parallelism switching strategy specialized at MoE training tasks. It also includes a hierarchical \op design to cope with the inter- and intra-node GPU topology for better efficiency. It is complementary with \sys.

FasterMoE~\cite{he2022fastermoe} 
proposes a roofline performance model to analyze the end-to-end performance 
of MoE training systems. Guided by this model, they propose a dynamic shadowing approach that pulls popular expert parameters instead of sending tokens to the experts. They also design a topology-aware expert selection strategy that 
relieves network congestion by sending tokens to experts with lower latency. 

\noindent\textbf{Communication acceleration in distributed training.}
Our community has proposed several communication schedulers for generic 
distributed training~\cite{MLSYS2019_84d9ee44, bytescheduler, bao2020preemptive, chen2020elastic, MLSYS2019_9b861925, MLSYS2020_43ec517d}. 
The objective is 
to better overlap the communication and computation operations in the backward 
pass and prioritize the communication of former layers over latter layers in 
the model. In \sys, we leverage the domain-specific insight that \op should be prioritized over \ar in MoE training, which is different from prior work. 
BytePS~\cite{byteps} proposes to reduce the communication traffic by utilizing the heterogeneous GPU/CPU resources in a training cluster. These acceleration techniques can be integrated into distributed MoE. \sys can also benefit from this idea, since more available bandwidth can be left to all-to-all operations.

%% file: main.bbl
\begin{thebibliography}{10}

\bibitem{ChatGPT}
{ChatGPT: Optimizing Language Models for Dialogue}.
\newblock \url{https://openai.com/blog/chatgpt/}.

\bibitem{deepspeed}
{DeepSpeed}.
\newblock \url{https://github.com/microsoft/DeepSpeed}.

\bibitem{wik8}
{Enwik8}.
\newblock \url{http://prize.hutter1.net/}.

\bibitem{nccl}
{NCCL}.
\newblock \url{https://github.com/NVIDIA/nccl }.

\bibitem{torchddp}
{PyTorch Distributed Data Parallel}.
\newblock \url{https://pytorch.org/docs/stable/notes/ddp.html }.

\bibitem{torchprofiler}
{PyTorch Profiler}.
\newblock \url{https://pytorch.org/blog/pytorch-profiler-1.9-released/}.

\bibitem{tutel}
{Tutel}.
\newblock \url{https://github.com/microsoft/tutel}.

\bibitem{wmt}
{WMT 19}.
\newblock
  \url{https://github.com/facebookresearch/fairseq/blob/main/examples/wmt19/README.md}.

\bibitem{artetxe2021efficient}
Mikel Artetxe, Shruti Bhosale, Naman Goyal, Todor Mihaylov, Myle Ott, Sam
  Shleifer, Xi~Victoria Lin, Jingfei Du, Srinivasan Iyer, Ramakanth Pasunuru,
  et~al.
\newblock Efficient large scale language modeling with mixtures of experts.
\newblock {\em arXiv preprint arXiv:2112.10684}, 2021.

\bibitem{FairScale2021}
Mandeep Baines, Shruti Bhosale, Vittorio Caggiano, Naman Goyal, Siddharth
  Goyal, Myle Ott, Benjamin Lefaudeux, Vitaliy Liptchinsky, Mike Rabbat, Sam
  Sheiffer, Anjali Sridhar, and Min Xu.
\newblock Fairscale: A general purpose modular pytorch library for high
  performance and large scale training.
\newblock \url{https://github.com/facebookresearch/fairscale}.

\bibitem{bao2020preemptive}
Yixin Bao, Yanghua Peng, Yangrui Chen, and Chuan Wu.
\newblock Preemptive all-reduce scheduling for expediting distributed dnn
  training.
\newblock In {\em IEEE INFOCOM}, 2020.

\bibitem{MLSYS2022_98dce83d}
Paul Barham, Aakanksha Chowdhery, Jeff Dean, Sanjay Ghemawat, Steven Hand,
  Daniel Hurt, Michael Isard, Hyeontaek Lim, Ruoming Pang, Sudip Roy, Brennan
  Saeta, Parker Schuh, Ryan Sepassi, Laurent Shafey, Chandu Thekkath, and
  Yonghui Wu.
\newblock Pathways: Asynchronous distributed dataflow for ml.
\newblock In {\em Proc.~MLSys}, 2022.

\bibitem{bengio2015conditional}
Emmanuel Bengio, Pierre-Luc Bacon, Joelle Pineau, and Doina Precup.
\newblock Conditional computation in neural networks for faster models.
\newblock {\em arXiv preprint arXiv:1511.06297}, 2015.

\bibitem{bengio2013estimating}
Yoshua Bengio, Nicholas L{\'e}onard, and Aaron Courville.
\newblock Estimating or propagating gradients through stochastic neurons for
  conditional computation.
\newblock {\em arXiv preprint arXiv:1308.3432}, 2013.

\bibitem{gpt-3}
Tom Brown, Benjamin Mann, Nick Ryder, Melanie Subbiah, Jared~D Kaplan, Prafulla
  Dhariwal, Arvind Neelakantan, Pranav Shyam, Girish Sastry, Amanda Askell,
  et~al.
\newblock Language models are few-shot learners.
\newblock {\em Advances in neural information processing systems},
  33:1877--1901, 2020.

\bibitem{chen2023sparse}
Tianlong Chen, Zhenyu Zhang, AJAY~KUMAR JAISWAL, Shiwei Liu, and Zhangyang
  Wang.
\newblock Sparse moe as the new dropout: Scaling dense and self-slimmable
  transformers.
\newblock In {\em Proc.~ICLR}, 2023.

\bibitem{chen2020elastic}
Yangrui Chen, Yanghua Peng, Yixin Bao, Chuan Wu, Yibo Zhu, and Chuanxiong Guo.
\newblock Elastic parameter server load distribution in deep learning clusters.
\newblock In {\em Proc.~ACM SoCC}, 2020.

\bibitem{chi2022representation}
Zewen Chi, Li~Dong, Shaohan Huang, Damai Dai, Shuming Ma, Barun Patra, Saksham
  Singhal, Payal Bajaj, Xia Song, Xian-Ling Mao, et~al.
\newblock On the representation collapse of sparse mixture of experts.
\newblock {\em Proc.~NeurIPS}, 2022.

\bibitem{MLSYS2019_9b861925}
Minsik Cho, Ulrich Finkler, David Kung, and Hillery Hunter.
\newblock Blueconnect: Decomposing all-reduce for deep learning on
  heterogeneous network hierarchy.
\newblock In {\em Proc.~MLSys}, 2019.

\bibitem{dai2019transformer}
Zihang Dai, Zhilin Yang, Yiming Yang, Jaime Carbonell, Quoc~V Le, and Ruslan
  Salakhutdinov.
\newblock {Transformer-XL: Attentive language models beyond a fixed-length
  context}.
\newblock {\em arXiv preprint arXiv:1901.02860}, 2019.

\bibitem{devlin2018bert}
Jacob Devlin, Ming-Wei Chang, Kenton Lee, and Kristina Toutanova.
\newblock Bert: Pre-training of deep bidirectional transformers for language
  understanding.
\newblock {\em arXiv preprint arXiv:1810.04805}, 2018.

\bibitem{du2021glam}
Nan Du, Yanping Huang, Andrew~M Dai, Simon Tong, Dmitry Lepikhin, Yuanzhong Xu,
  Maxim Krikun, Yanqi Zhou, Adams~Wei Yu, Orhan Firat, Barret Zoph, Liam Fedus,
  Maarten~P Bosma, Zongwei Zhou, Tao Wang, Emma Wang, Kellie Webster, Marie
  Pellat, Kevin Robinson, Kathleen Meier-Hellstern, Toju Duke, Lucas Dixon, Kun
  Zhang, Quoc Le, Yonghui Wu, Zhifeng Chen, and Claire Cui.
\newblock {GL}a{M}: Efficient scaling of language models with
  mixture-of-experts.
\newblock In {\em Proceedings of Machine Learning Research}, 2022.

\bibitem{fedus2021switch}
William Fedus, Barret Zoph, and Noam Shazeer.
\newblock {Switch transformers: Scaling to trillion parameter models with
  simple and efficient sparsity}.
\newblock {\em arXiv preprint arXiv:2101.03961}, 2021.

\bibitem{MLSYS2019_84d9ee44}
Sayed~Hadi Hashemi, Sangeetha Abdu~Jyothi, and Roy Campbell.
\newblock Tictac: Accelerating distributed deep learning with communication
  scheduling.
\newblock In A.~Talwalkar, V.~Smith, and M.~Zaharia, editors, {\em
  Proc.~MLSys}, 2019.

\bibitem{he2022fastermoe}
Jiaao He, Jidong Zhai, Tiago Antunes, Haojie Wang, Fuwen Luo, Shangfeng Shi,
  and Qin Li.
\newblock {FasterMoE: modeling and optimizing training of large-scale dynamic
  pre-trained models}.
\newblock In {\em Proc.~ACM SIGPLAN PPoPP}, pages 120--134, 2022.

\bibitem{jia2020whale}
Xianyan Jia, Le~Jiang, Ang Wang, Jie Zhang, Xinyuan Li, Wencong Xiao, Yong Li,
  Zhen Zheng, Xiaoyong Liu, Wei Lin, et~al.
\newblock Whale: Scaling deep learning model training to the trillions.
\newblock {\em arXiv preprint arXiv:2011.09208}, 2020.

\bibitem{jia2019beyond}
Zhihao Jia, Matei Zaharia, and Alex Aiken.
\newblock Beyond data and model parallelism for deep neural networks.
\newblock {\em Proc.~MLSys}, 2019.

\bibitem{byteps}
Yimin Jiang, Yibo Zhu, Chang Lan, Bairen Yi, Yong Cui, and Chuanxiong Guo.
\newblock A unified architecture for accelerating distributed {DNN} training in
  heterogeneous {GPU/CPU} clusters.
\newblock In {\em Proc.~USENIX OSDI}, pages 463--479, 2020.

\bibitem{kaplan2020scaling}
Jared Kaplan, Sam McCandlish, Tom Henighan, Tom~B Brown, Benjamin Chess, Rewon
  Child, Scott Gray, Alec Radford, Jeffrey Wu, and Dario Amodei.
\newblock Scaling laws for neural language models.
\newblock {\em arXiv preprint arXiv:2001.08361}, 2020.

\bibitem{komatsuzaki2023sparse}
Aran Komatsuzaki, Joan Puigcerver, James Lee-Thorp, Carlos~Riquelme Ruiz, Basil
  Mustafa, Joshua Ainslie, Yi~Tay, Mostafa Dehghani, and Neil Houlsby.
\newblock Sparse upcycling: Training mixture-of-experts from dense checkpoints.
\newblock In {\em Proc.~ICLR}, 2023.

\bibitem{lepikhin2020gshard}
Dmitry Lepikhin, HyoukJoong Lee, Yuanzhong Xu, Dehao Chen, Orhan Firat, Yanping
  Huang, Maxim Krikun, Noam Shazeer, and Zhifeng Chen.
\newblock {Gshard: Scaling giant models with conditional computation and
  automatic sharding}.
\newblock {\em arXiv preprint arXiv:2006.16668}, 2020.

\bibitem{pmlr-v139-lewis21a}
Mike Lewis, Shruti Bhosale, Tim Dettmers, Naman Goyal, and Luke Zettlemoyer.
\newblock Base layers: Simplifying training of large, sparse models.
\newblock In {\em Proc.~USENIX ICML}, 2021.

\bibitem{liang2022mvit}
Hanxue Liang, Zhiwen Fan, Rishov Sarkar, Ziyu Jiang, Tianlong Chen, Kai Zou,
  Yu~Cheng, Cong Hao, and Zhangyang Wang.
\newblock M{\textthreesuperior}vit: Mixture-of-experts vision transformer for
  efficient multi-task learning with model-accelerator co-design.
\newblock In {\em Proc.~NeurIPS}, 2022.

\bibitem{imdb}
Andrew~L. Maas, Raymond~E. Daly, Peter~T. Pham, Dan Huang, Andrew~Y. Ng, and
  Christopher Potts.
\newblock Learning word vectors for sentiment analysis.
\newblock In {\em Proc.~ACL}, 2011.

\bibitem{paws2019naacl}
Ibrahim Naji.
\newblock {TSATC: Twitter Sentiment Analysis Training Corpus}.
\newblock In {\em thinknook}, 2012.

\bibitem{OPRV13}
Kay Ousterhout, Aurojit Panda, Joshua Rosen, Shivaram Venkataraman, Reynold
  Xin, Sylvia Ratnasamy, Scott Shenker, and Ion Stoica.
\newblock {The Case for Tiny Tasks in Compute Clusters}.
\newblock In {\em Proc.~USENIX HotOS}, 2013.

\bibitem{bytescheduler}
Yanghua Peng, Yibo Zhu, Yangrui Chen, Yixin Bao, Bairen Yi, Chang Lan, Chuan
  Wu, and Chuanxiong Guo.
\newblock A generic communication scheduler for distributed dnn training
  acceleration.
\newblock In {\em Proc.~ACM SOSP}, 2019.

\bibitem{msMoE}
Andrey Proskurin.
\newblock {DeepSpeed: Advancing MoE inference and training to power
  next-generation AI scale}.
\newblock
  \url{https://www.microsoft.com/en-us/research/blog/deepspeed-advancing-moe-inference-and-training-to-power-next-generation-ai-scale}.

\bibitem{radford2019language}
Alec Radford, Jeff Wu, Rewon Child, David Luan, Dario Amodei, and Ilya
  Sutskever.
\newblock Language models are unsupervised multitask learners.
\newblock {\em OpenAI blog}, 2019.

\bibitem{2020t5}
Colin Raffel, Noam Shazeer, Adam Roberts, Katherine Lee, Sharan Narang, Michael
  Matena, Yanqi Zhou, Wei Li, and Peter~J. Liu.
\newblock Exploring the limits of transfer learning with a unified text-to-text
  transformer.
\newblock {\em Journal of Machine Learning Research}, 2020.

\bibitem{rajbhandari2022deepspeedmoe}
Samyam Rajbhandari, Conglong Li, Zhewei Yao, Minjia Zhang, Reza~Yazdani
  Aminabadi, Ammar~Ahmad Awan, Jeff Rasley, and Yuxiong He.
\newblock {DeepSpeed-MoE: Advancing Mixture-of-Experts Inference and Training
  to Power Next-Generation AI Scale}.
\newblock {\em arXiv preprint arXiv:2201.05596}, 2022.

\bibitem{ren2021zero}
Jie Ren, Samyam Rajbhandari, Reza~Yazdani Aminabadi, Olatunji Ruwase, Shuangyan
  Yang, Minjia Zhang, Dong Li, and Yuxiong He.
\newblock {ZeRO-Offload: Democratizing Billion-Scale Model Training}.
\newblock In {\em Proc.~USENIX ATC}, 2021.

\bibitem{shazeer2018mesh}
Noam Shazeer, Youlong Cheng, Niki Parmar, Dustin Tran, Ashish Vaswani, Penporn
  Koanantakool, Peter Hawkins, HyoukJoong Lee, Mingsheng Hong, Cliff Young,
  Ryan Sepassi, and Blake Hechtman.
\newblock {Mesh-TensorFlow}: Deep learning for supercomputers.
\newblock In {\em Proc.~ACM NeurIPS}, 2018.

\bibitem{shazeer2017outrageously}
Noam Shazeer, Azalia Mirhoseini, Krzysztof Maziarz, Andy Davis, Quoc Le,
  Geoffrey Hinton, and Jeff Dean.
\newblock {Outrageously large neural networks: The sparsely-gated
  mixture-of-experts layer}.
\newblock {\em arXiv preprint arXiv:1701.06538}, 2017.

\bibitem{shen2022se}
Liang Shen, Zhihua Wu, WeiBao Gong, Hongxiang Hao, Yangfan Bai, HuaChao Wu,
  Xinxuan Wu, Haoyi Xiong, Dianhai Yu, and Yanjun Ma.
\newblock Se-moe: A scalable and efficient mixture-of-experts distributed
  training and inference system.
\newblock {\em arXiv preprint arXiv:2205.10034}, 2022.

\bibitem{shoeybi2019megatron}
Mohammad Shoeybi, Mostofa Patwary, Raul Puri, Patrick LeGresley, Jared Casper,
  and Bryan Catanzaro.
\newblock Megatron-lm: Training multi-billion parameter language models using
  model parallelism.
\newblock {\em arXiv preprint arXiv:1909.08053}, 2019.

\bibitem{vaswani2017attention}
Ashish Vaswani, Noam Shazeer, Niki Parmar, Jakob Uszkoreit, Llion Jones,
  Aidan~N Gomez, {\L}ukasz Kaiser, and Illia Polosukhin.
\newblock Attention is all you need.
\newblock {\em Advances in neural information processing systems}, 30, 2017.

\bibitem{MLSYS2020_43ec517d}
Guanhua Wang, Shivaram Venkataraman, Amar Phanishayee, Nikhil Devanur, Jorgen
  Thelin, and Ion Stoica.
\newblock Blink: Fast and generic collectives for distributed ml.
\newblock In {\em Proc.~MLSys}, 2020.

\bibitem{wolf-etal-2020-transformers}
Thomas Wolf, Lysandre Debut, Victor Sanh, Julien Chaumond, Clement Delangue,
  Anthony Moi, Pierric Cistac, Tim Rault, Rémi Louf, Morgan Funtowicz, Joe
  Davison, Sam Shleifer, Patrick von Platen, Clara Ma, Yacine Jernite, Julien
  Plu, Canwen Xu, Teven~Le Scao, Sylvain Gugger, Mariama Drame, Quentin Lhoest,
  and Alexander~M. Rush.
\newblock Transformers: State-of-the-art natural language processing.
\newblock In {\em Proc.~EMNLP}, 2020.

\bibitem{yao2022random}
Zhewei Yao, Xiaoxia Wu, Conglong Li, Connor Holmes, Minjia Zhang, Cheng Li, and
  Yuxiong He.
\newblock Random-ltd: Random and layerwise token dropping brings efficient
  training for large-scale transformers.
\newblock {\em arXiv preprint arXiv:2211.11586}, 2022.

\bibitem{zhang2017poseidon}
Hao Zhang, Zeyu Zheng, Shizhen Xu, Wei Dai, Qirong Ho, Xiaodan Liang, Zhiting
  Hu, Jinliang Wei, Pengtao Xie, and Eric~P Xing.
\newblock {Poseidon: An efficient communication architecture for distributed
  deep learning on GPU clusters}.
\newblock In {\em Proc.~USENIX ATC}, 2017.

\bibitem{zheng2022alpa}
Lianmin Zheng, Zhuohan Li, Hao Zhang, Yonghao Zhuang, Zhifeng Chen, Yanping
  Huang, Yida Wang, Yuanzhong Xu, Danyang Zhuo, Eric~P Xing, et~al.
\newblock Alpa: Automating inter-and intra-operator parallelism for distributed
  deep learning.
\newblock In {\em Proc.~USENIX OSDI}, 2022.

\bibitem{zhou2022mixture}
Yanqi Zhou, Tao Lei, Hanxiao Liu, Nan Du, Yanping Huang, Vincent Zhao, Andrew~M
  Dai, Quoc~V Le, James Laudon, et~al.
\newblock Mixture-of-experts with expert choice routing.
\newblock {\em Proc.~NeurIPS}, 2022.

\bibitem{zoph2022designing}
Barret Zoph, Irwan Bello, Sameer Kumar, Nan Du, Yanping Huang, Jeff Dean, Noam
  Shazeer, and William Fedus.
\newblock {Designing Effective Sparse Expert Models}.
\newblock {\em arXiv preprint arXiv:2202.08906}, 2022.

\bibitem{zuo2021taming}
Simiao Zuo, Xiaodong Liu, Jian Jiao, Young~Jin Kim, Hany Hassan, Ruofei Zhang,
  Tuo Zhao, and Jianfeng Gao.
\newblock Taming sparsely activated transformer with stochastic experts.
\newblock {\em arXiv preprint arXiv:2110.04260}, 2021.

\end{thebibliography}
